\documentclass[fleqn,usenatbib]{mnras}
\usepackage{amsmath}    
\usepackage{newtxtext,newtxmath}
\usepackage[T1]{fontenc}
\usepackage{ae,aecompl}
\usepackage{graphicx}	
\usepackage{geometry}
\usepackage{float}
\usepackage{caption}
\usepackage{subcaption}
\usepackage{dblfloatfix}
\usepackage{url}
\usepackage{xcolor,colortbl}

\newcommand{\hmpc}{\,h^{-1}\,{\rm Mpc}}
\newcommand{\hmpcc}{\,h^{3}\,{\rm Mpc}^{-3}}
\newcommand{\hmsun}{\,h^{-1}\,{\rm M}_\odot}
\newcommand{\vvir}{{\rm V_{vir}}}

\newcommand{\vmax}{{\rm V_{max}}}
\newcommand{\vpeak}{{\rm V_{peak}}}

\newcommand{\Mh}{M_{\rm h}}

\newcommand{\Mstar}{M_*}
\newcommand{\Msun}{{\rm M}_\odot}

\title{Dissecting and modelling galaxy assembly bias}

\author[X. Xu et al.]{
Xiaoju Xu,$^{1}$\thanks{E-mail: xiaoju.xu@case.edu}
Idit Zehavi,$^{1}$\thanks{E-mail: idit.zehavi@case.edu}
and Sergio Contreras$^{2}$\\
$^{1}$Department of Physics, Case Western Reserve University, 10900 Euclid Avenue, Cleveland, OH 44106, USA\\
$^{2}$Donostia International Physics Center (DIPC), Manuel Lardizabal Ibilbidea, 4, 20018 Donostia, Gipuzkoa, Spain.}

\date{Accepted XXX. Received YYY; in original form ZZZ}

\pubyear{2020}

\begin{document}
\label{firstpage}
\pagerange{\pageref{firstpage}--\pageref{lastpage}}
\maketitle

\begin{abstract}
Understanding the galaxy-halo connection is fundamental for contemporary models of galaxy clustering. The extent to which the haloes' assembly history and environment impact galaxy clustering (a.k.a. galaxy assembly bias; GAB), remains a complex and challenging problem. Using a semi-analytic galaxy formation model,  we study the individual contributions of different secondary halo properties to the GAB signal. These are obtained by comparing the clustering of stellar-mass selected samples to that of shuffled samples where the galaxies are randomly reassigned to haloes of fixed mass and a specified secondary halo property. We explore a large range of internal halo properties and environmental measures. We find that commonly used properties like halo age or concentration amount to only 20-30 per cent of the signal, while the smoothed matter density or the tidal anisotropy can account for the full level of GAB (though care should be given to the specific definition). For the ``successful'' measures, we examine the occupancy variations and the associated changes in the halo occupation function parameters. These are used to create mock catalogues that reproduce the full level of GAB. Finally, we propose a practical modification of the standard halo occupation distribution model, which can be tuned to any level of assembly bias.  Fitting the parameters to our semi-analytic model, we demonstrate that the corresponding mock catalogue recovers the target level of GAB as well as the occupancy variations. Our results enable producing realistic mock catalogues and directly inform theoretical modelling of assembly bias and attempts to detect it in the Universe.
  
\end{abstract}


\section{Introduction}
\label{sec:intro}

In the standard picture of hierarchical structure formation, galaxies reside in dark matter haloes \citep{White1978}.  The formation and evolution of the haloes are dominated by gravity, while the formation of the galaxies and their relation to the dark matter haloes are more complex and depend on the detailed physical processes. Galaxy clustering is a fundamental observable that can be used to constrain both cosmological parameters and galaxy formation physics.  It is crucial to have a detailed understanding of the connection between galaxies and their host haloes (see \citealt{Wechsler2018} for a review), if we are to optimally use galaxies as a cosmological probe. This is particularly essential with the advent of large galaxy surveys aimed at measuring galaxy clustering with unparalleled accuracy.

As the basis for galaxy clustering, halo clustering can be modelled by analytical theory and $N$-body simulations that trace the formation and evolution of dark matter haloes under the gravitational influence (see \citealt{Cooray2002} for a review). In traditional analytical models, dark matter halo clustering is modelled as a function of halo mass \citep{Press74,Bond91,Sheth99,Sheth01,Tinker08b}. However, with the help of high-resolution cosmological $N$-body simulations \citep{Springel2005,Prada2012}, numerous studies have shown that halo clustering depends also on properties related to the assembly history of the haloes, such as formation time or concentration (e.g. \citealt{Gao05,Wechsler06,Gao07,Sato2018}), as well as the tidal environment of the haloes \citep{Paranjape2018a,Ramakrishnan2019,Mansfield2020,Ramakrishnan2020}. The dependence of halo clustering on properties beyond halo mass has commonly been referred to as {\it halo assembly bias} (HAB hereafter). For example, haloes that formed earlier, or with higher concentration, cluster more strongly than late-formed or low concentration haloes of the same mass. Haloes in higher density regions or with stronger tidal anisotropy also have higher bias than the ones in lower density regions or lower tidal anisotropy. Although it is difficult, efforts have been made to incorporate such secondary effects on halo bias into analytical models \citep{Dalal08,Shi2018}. 

Beyond halo clustering, the halo occupation function, which describes the average number of galaxies of a given type as a function of halo mass, is a key component in determining galaxy clustering. In particular, it provides the basis for the Halo Occupation Distribution (HOD) framework (e.g. \citealt{Peacock2000,Scoccimarro01,Berlind02,Zheng2005,Zheng07}), which characterizes the galaxy-halo connection statistically at the level of individual haloes. This approach has been very powerful in explaining the shape of the galaxy correlation function, its evolution, and dependence on galaxy properties (e.g. \citealt{Zehavi2004,Zehavi2005,Zehavi2011,Zheng2007,Coupon12,Contreras2017}).  It has also become a very popular tool for populating haloes in large simulations in order to produce realistic mock catalogues (e.g. \citealt{manera15,zheng16,smith17,AEMULUS}), increasingly important for the planning and analysis of galaxy surveys.

In the standard HOD approach, or in Conditional Luminosity Function modelling \citep{Bosch03,Bosch13,Yang03,Yang08}, the halo occupation function is considered to depend on only halo mass. However, if galaxy properties closely correlate with the halo formation history, we expect the galaxy content of haloes to also depend on the secondary halo properties, leading to a dependence on the large-scale environment and consequently impacting galaxy clustering. Such {\it occupancy variations} (OV hereafter) have been explored in detail in both semi-analytic galaxy formation models and hydrodynamical simulations \citep{Zhu06,Zehavi2018,Zehavi2019,Artale2018,Contreras2019,Bose2019},  finding distinct variations with halo formation time and concentration, and more subtle features with environment. \citet{Montero2020} and \citet{Xu2020} furthermore study the explicit correlation of galaxy properties with halo properties in such galaxy formation models.
Neglecting these effects can have direct implications for interpreting galaxy clustering using the HOD framework \citep{Pujol14,Zentner2014,Lange19}.  To mitigate that, some studies aim to generalise the HOD to incorporate the dependence on secondary halo properties or environment (e.g. \citealt{Paranjape2015,Hearin2016,McEwen2018,Hadzhiyska2020b}). Attempts have also been made to absorb the secondary effects with primary halo properties other than halo mass \citep{Dragomir2018,Zehavi2019}, with mixed results. 

The combined effect of HAB and the OV lead to a potential change of the amplitude of galaxy clustering on large scales. This imprint of assembly bias on the galaxy distribution is broadly referred to as {\it galaxy assembly bias} (GAB henceforth).  It has commonly been explored in simulations by comparing the large-scale correlation function of a specified galaxy sample to that of a shuffled galaxy sample, where the galaxies are randomly reassigned among haloes of the same mass (e.g. \citealt{Croton2007,Zu08,Chaves2016}). The shuffling eliminates the connection between galaxies and any secondary halo property, effectively removing the OV. While HAB and OV both depend on the specific secondary property studied, the GAB signature is the full (net) effect resulting from all secondary properties combined.
For HAB, it has proven quite challenging to derive a specific parameter or combination thereof that captures all of the measured trends (e.g. \citealt{mao17,Villarreal17,XuX18,Salcedo18,Han2019}). While many assembly bias studies focus on halo age and concentration as the main properties, other studies stress the key role played by environment \citep{McEwen2018,Shi2018,Han2019} and recent claims suggest that tidal anisotropy is a primary indicator of assembly bias \citep{Ramakrishnan2019,Mansfield2020}.

In what follows, we extend the shuffling methodology to systematically explore the individual contributions to GAB.  We use here the \citet{Guo2011} semi-analytic galaxy formation model applied to the Millennium simulation \citep{Springel2005}, and analyse stellar-mass selected galaxy samples corresponding to different number densities. We build on the work of \citet{Croton2007} who investigated the effects of halo age and concentration on GAB.  We expand on this work and examine a large variety of internal halo properties as well as environmental and tidal anisotropy measures. Interestingly, we find that most of the secondary halo properties produce only a small fraction of the GAB signal, while environmental and tidal anisotropy measures are better able to ``capture'' the full effect.  We investigate the OV and the changes in the standard HOD parameters, and utilize them to produce mock catalogues that incorporate the GAB effect. Our results are comparable to those of \citet{Hadzhiyska2020a} who study the impact of secondary halo and environment properties on GAB in Illustris TNG hydrodynamical simulation \citep{Nelson2019}, using an abundance-matching inspired method, where the GAB contained in a secondary property is boosted to the maximum.
In addition to elucidating the individual contributions of different halo and environmental properties to the GAB signal, we also explore the possibility of modifying the traditional HOD model using the most important secondary properties. We propose a seven-parameter HOD model, in a similar spirit of some recent studies \citep{McEwen2018,Wibking2019,Salcedo2020}, which can be extremely useful for incorporating assembly bias into mock catalogues.

Though GAB has been studied extensively in different galaxy formation models and simulations, it is difficult to infer it directly from observations,  and there is no clear consensus on its existence in the real Universe. Some detections have been suggested \citep{cooper10,wang13,Hearin15,miyatake16,montero17,Ferreras19,Obuljen2020}  while others indicate the impact of assembly bias to be small \citep{Abbas06,Blanton07,Tinker08a,Lin2016,Zu2016,Walsh19} and that previous claims were plagued by systematics \citep[e.g.][]{Campbell15,Zu2017,Sin17,tinker17a,Lacerna2018,Sunayama19}. An important upcoming application is to use our modified HOD to determine the level of assembly bias in the Universe.

The structure of the paper is organized as follows: In \S~\ref{sec:sim}, the $N$-body simulation and semi-analytic galaxy formation model are introduced, as well as the halo and environment properties used in this work. In \S~\ref{sec:gab}, we present the result of GAB dependence on secondary properties. We show the occupancy variation caused by specific environment properties in \S~\ref{sec:hodmock}, and build mock catalogues based on the occupancy variation and five-parameter HOD. In \S~\ref{sec:edhod}, we present a modified seven-parameter HOD that incorporates assembly bias and construct mock catalogues based on it. We conclude in \S~\ref{sec:summary}. Appendix~\ref{sec:appendix} presents additional results relating to the tidal anisotropy, and Appendix~\ref{sec:appendix2} includes more results of our modified HOD.   

\section{Simulated data}
\label{sec:sim}

\subsection{Numerical simulation and galaxy formation model}
\label{subsec:simsam}

We employ in this work the Millennium simulation. The Millennium simulation \citep{Springel2005} is a dark matter only $N$-body simulation of $2160^3$ dark matter particles of mass $8.6\times10^8 \hmsun$. The simulated volume is a periodic comoving box of 500 $h^{-1}{\rm Mpc}$ on aside. The simulation was run using GADGET-2 \citep{Springel2005} from $z=127$ to $z=0$, and outputs 64 snapshots at different redshifts. The simulation assumes a $\Lambda{\rm CDM}$ cosmology for which the parameters are $\Omega_{\rm m}=0.25$, $\Omega_{\rm b}=0.045$, $h=0.73$, $\sigma_8=0.9$, and $n_s$=1. At each snapshot, the haloes are identified using a friends-of-friends algorithm \citep{Davis1985} for structures above 20 particles. The subhaloes and the merger trees are constructed with the \texttt{SUBFIND} algorithm \citep{Springel2001}, which are later used to populate the simulation with galaxies. The large volume of this simulation allows us to have good statistics for our analyses, and to robustly measure galaxy clustering on large scales. The high resolution allows us to also reliably explore the galaxy-halo connection.

We build our galaxy catalogues using the semi-analytic model of \citet{Guo2011}. Semi-analytic models (SAMs) aim to follow the main physical processes involved in galaxy formation and evolution in a cosmological framework. These processes include prescriptions for star formation, gas cooling, supernovae and active galactic nuclei feedback, chemical evolution, and galaxy mergers. The SAMs use the merger trees from $N$-body simulations or an extended Press-Schechter formalism as a basis to form and evolve the galaxies. This has become a popular technique to study galaxy formation for its capacity to track galaxies in very large cosmological volumes (up to a few ${\rm Gpc^3}$) with relatively low computational power and high predictive power (e.g. \citealt{Henriques2015,Stevens2018,Lacey2016,Croton2016,Lagos2018}). For more extended reviews on SAMs, see \citet{Baugh2006} and \citet{Benson2010}.

The \citet{Guo2011} model is a flavour of L-Galaxies, the semi-analytic code from the Munich group \citep{Springel2001,DeLucia2004,Croton2006,Bertone2007,Guo2013,Henriques2020}. The model is calibrated by fitting to observational data, such as the stellar mass function and luminosity function at low redshift, as well as the relation between black hole mass and bulge mass. The outputs of the models are publicly available in the Millennium database webpage\footnote{\url{http://gavo.mpa-garching.mpg.de/Millennium/}} via SQL protocol.
We chose this particular SAM and dark matter simulation because of the number of related works done with this simulation (e.g. \citealt{Gao05,Gao07,Croton2007,Li2008,Zehavi2018}) and the large amount of available properties for the haloes, subhaloes and the environment of the simulation.
We focus here on three galaxy samples with different number densities, ranked by the stellar mass of the galaxies. The three number densities are $n_1=0.00316 \hmpcc$, $n_2=0.01 \hmpcc$, and $n_3=0.0316 \hmpcc$, which correspond to stellar-mass thresholds of $3.88\times10^{10} \hmsun$, $1.42\times10^{10} \hmsun$, and $0.185\times10^{10} \hmsun$, respectively. The samples are approximately evenly spaced in logarithmic number density, and follow the choices made in \citet{Zehavi2018}. 

\subsection{Halo properties and environmental measures}
\label{subsec:properties}

In this work, we investigate the individual impact of different secondary properties on GAB. The secondary properties used can be separated into two categories, internal halo properties and ``external'' measures of the environment of the haloes. The internal halo properties we utilize are: \newline 

\begin{itemize}
    \item[(1)] $a_{\rm 0.5}$, the scale factor at which the halo reaches for the first time half of its current mass.  This is commonly referred to as the halo age or formation time;
    \item[(2)] $c$, halo concentration parameter. We use the definition of $c=\vmax/\vvir$, where $\vmax$ is the maximum circular velocity of the halo and $\vvir$ is the virial velocity (e.g. \citealt{Bullock01});
    \item[(3)] $j$, the specific angular momentum of halo, i.e. the ratio of total angular momentum to the halo mass;
    \item[(4)] $\vmax$, the maximum circular velocity of the halo  
    \item[(5)] $\vpeak$, the peak value of $\vmax$ over the halo's accretion history;
    \item[(6)] $a_{\rm vpeak}$, the cosmic scale factor when $\vmax (a_{\rm vpeak}) = \vpeak$;
    \item[(7)] $a_{\rm first}$, the scale factor of the first major merger on the main branch of the halo merger tree. We define a major merger when the mass ratio of two progenitors is larger than 1/3;
    \item[(8)] $a_{\rm last}$, the scale factor of the last major merger on the main branch of the halo merger tree;
    \item[(9)] $n_{\rm sub}$, the total number of subhaloes associated with the halo as identified by \texttt{SUBFIND};
    \item[(10)] $n_{\rm sat}$, the total number of satellite galaxies residing in the halo as assigned by the SAM.   
\end{itemize}
The external environment measures we consider include the following categories: 
\begin{itemize}
    \item[(1)] $\delta_{1.25}$/$\delta_{2.5}$/$\delta_{5}$/$\delta_{10}$, the dark matter density smoothed with a Gaussian filter with a smoothing scale of $1.25$, $2.5$, $5$, and $10 \hmpc$, respectively. The smoothing is done by first measuring the counts-in-cell dark matter particle density on a $256^3$ grid, and then multiplying with a Gaussian kernel in Fourier space. The Millennium data provides these values at grid points in the simulation volume, and we perform a 3D interpolation between these points to obtain the density at the position of each halo; 
    \item[(2)] $\alpha_{\rm n,R}$, the tidal anisotropy measured with a Gaussian smoothing scale R. The potential field $\phi$ is obtained by solving the Poisson equation  $\nabla^2 \phi=-4\pi G \rho$, from which we calculate the tidal tensor $T_{ij}=\frac{\partial^2 \phi}{\partial x_i^2 \partial x_j^2}$. The tidal anisotropy parameter \citep{Paranjape2018a,Alam2019} is defined as
      \begin{equation}
      \label{eq:alpha}  
      \alpha_{\rm n,R}\equiv\sqrt{q_R^2}/(1+\delta_R)^{n}\,,
      \end{equation}
      where the tidal torque $q_R^2$ is
      \begin{equation}
      \label{eq:alpha2}  
      q_R^2=\frac{1}{2}[(\lambda_3-\lambda_2)^2+(\lambda_3-\lambda_1)^2+(\lambda_2-\lambda_1)^2]\,,
      \end{equation}
      with $\lambda_1$,$\lambda_2$,$\lambda_3$ the eigenvalues of the tidal tensor \citep{Heavens1988,Catelan1996}. The tidal anisotropy is typically measured numerically from the density field, but it can also be calculated analytically \citep{Paranjape2020}. For the normalisation power $n$, we focus on $n$=1, 0.55, and 0.3, as discussed below;
    \item[(3)] $\alpha_{\rm type}$, the cosmic web type defined by the number of positive eigenvalues of the tidal tensor, characterizing a void, sheet, filament, and node environments, corresponding to zero, one, two, and three positive eigenvalues, respectively \citep{Hahn2007};
     \item[(4)] $r_{\rm 10}$, the minimum value of the ratio of the distance to nearby massive haloes, more massive than 10 times that of the halo considered, and their virial radius.  This variable is inversely proportional to the cube root of the tidal force produced, and thus is a simplistic measure of the largest tidal force imparted by a neighboring halo.
\end{itemize}

\section{Impact of secondary properties on GAB}
\label{sec:gab}
\subsection{Shuffling methodology}
\label{subsec:shuffling}

To study the impact of assembly bias on galaxy clustering in simulated datasets, it is standard practice to compare the correlation function of the original sample with that of a shuffled galaxy sample, where the galaxy content of haloes is randomly reassigned among haloes of the same mass \citep{Croton2007}.  In the first stage, we follow this methodology in order to measure the total amount of GAB in our samples. More specifically, we shuffle the central galaxies among haloes of the same mass bin. The satellite galaxies are moved together with their original central galaxy maintaining the same relative distribution, and thus preserving the same one-halo contribution to the correlation function. The halo mass bin we adopt is 0.1 dex below  $\log[\Mh/(\hmsun)]=14.6$, and 0.2 dex and 0.6 dex for the following two bins (due to the paucity of high-mass haloes). By virtue of the random reassignment, the galaxy content of any give halo may shift to either a halo that was previously occupied by galaxies in the sample or to a halo of that mass that was previously unoccupied. (The central galaxy is placed at the position of the former occupant or at the location of the most bound particle in the halo, in the latter case.) The shuffling procedure effectively removes the connection of the galaxy population to the assembly history of the haloes, and eliminates the dependence on any secondary properties other than halo mass.

\begin{figure*}
	\centering
	\begin{subfigure}[h]{0.48\textwidth}
	\includegraphics[width=\textwidth]{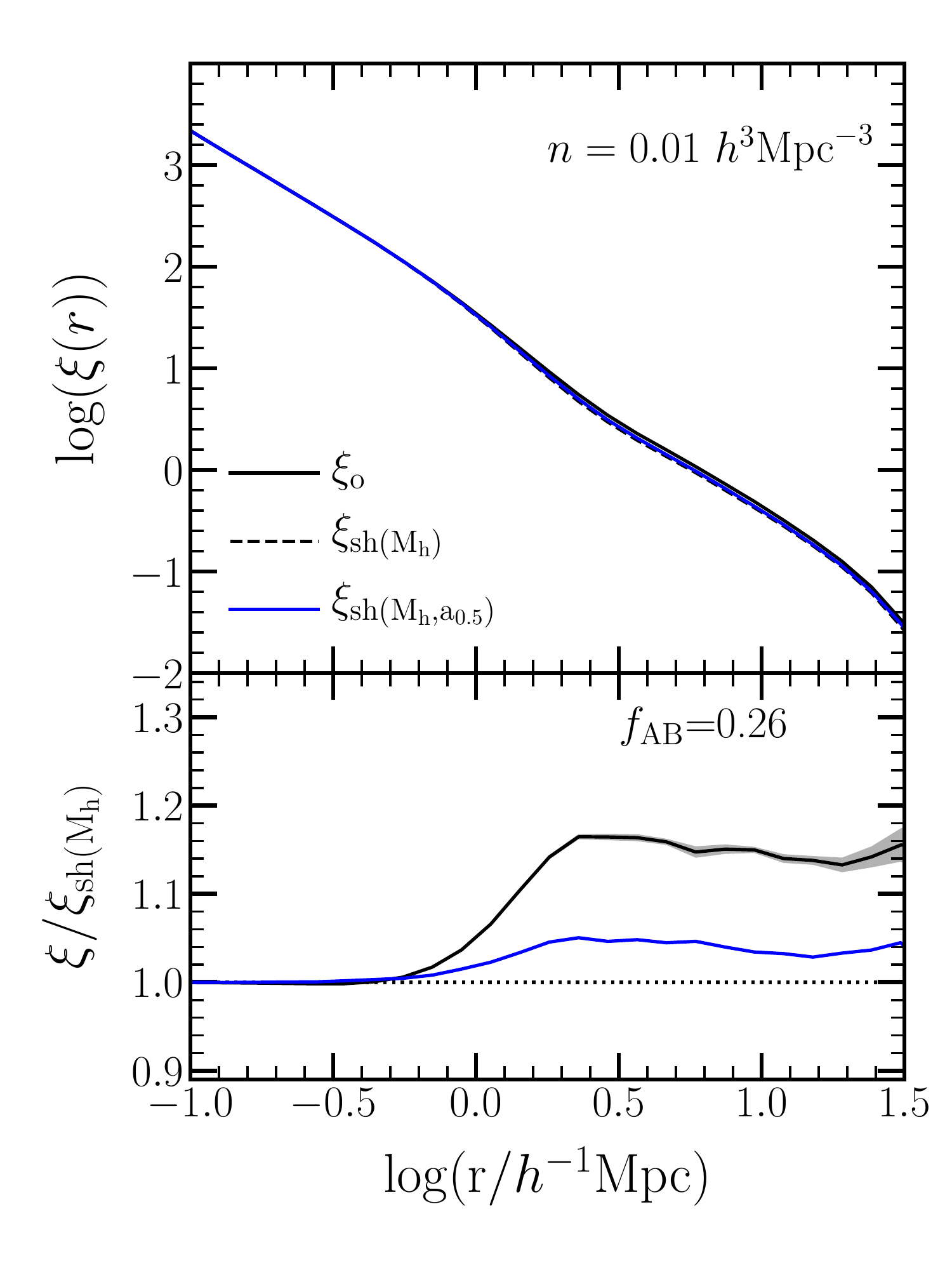}
	\end{subfigure}
	\hfill
	\begin{subfigure}[h]{0.48\textwidth}
        \includegraphics[width=\textwidth]{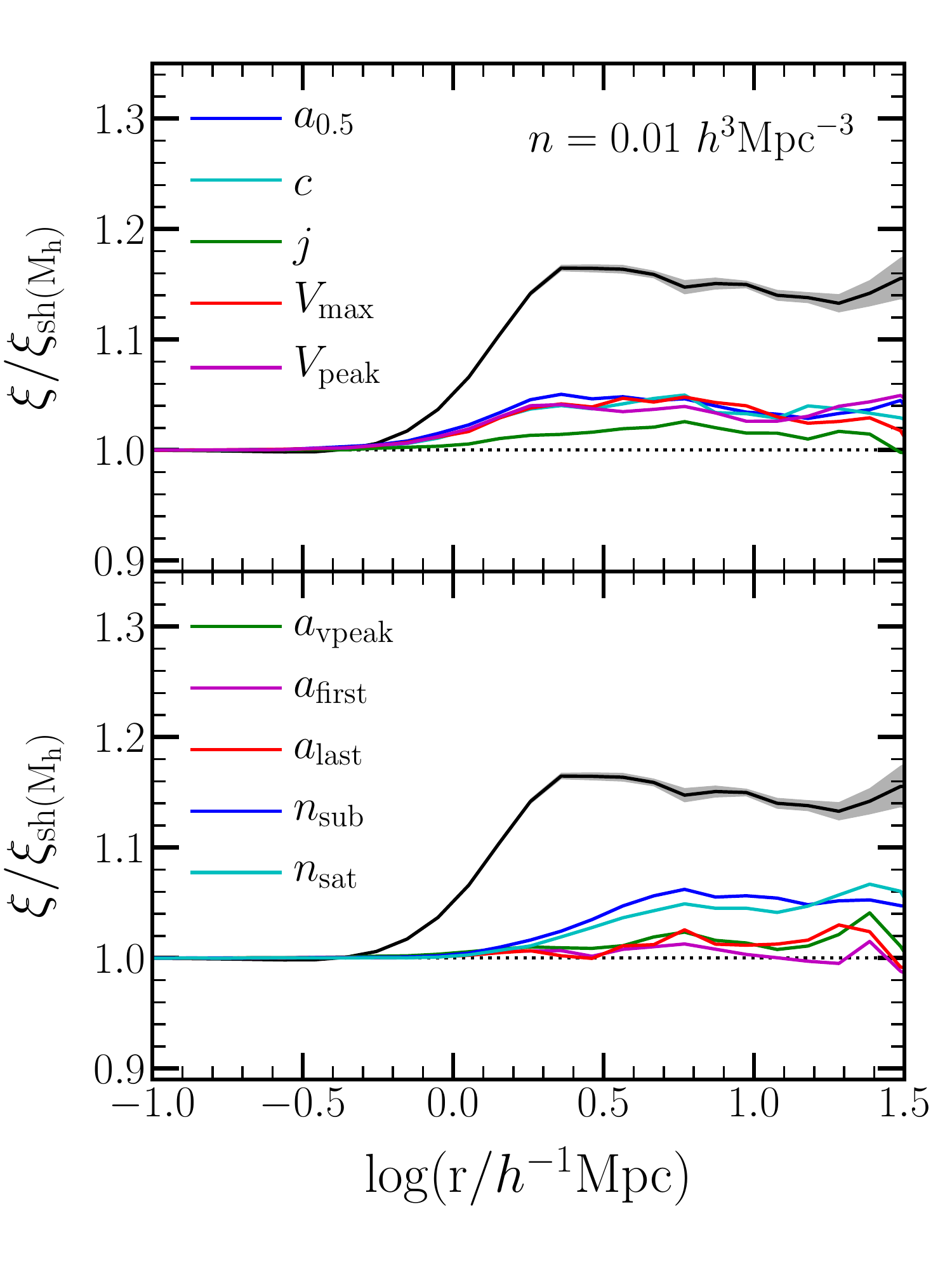}
	\end{subfigure}
	\hfill
\caption{
Auto-correlation functions for the galaxy sample with number density $n=0.01 \hmpcc$, exhibiting the impact of galaxy assembly bias. Left: top panel shows the correlation function of the original galaxy sample $\xi_{\rm o}$ (black solid line), the shuffled-by-mass sample $\xi_{\rm sh(\Mh)}$ (black dashed line), and the sample obtained by shuffling while fixing both mass and halo age $\xi_{\rm sh(\Mh,a_{0.5})}$ (blue solid line). In the bottom panel, the black solid shows the ratio $\xi_{\rm o}/\xi_{\rm sh(\Mh)}$ which presents the full GAB in the SAM. The blue solid is the ratio of $\xi_{\rm sh(\Mh,a_{0.5})}/\xi_{\rm sh(\Mh)}$ which is the GAB attributed to halo age (see text). A ratio of 1 is indicated by the dotted line. Also marked is the corresponding value of $f_{\rm AB}$, indicating the fraction of the full GAB that can be recovered by the secondary property (here halo age). Right: the same as the bottom panel on the left-hand side, but now for all internal halo properties considered,  namely the GAB contained by the secondary properties. In addition to $a_{0.5}$, these include $c$, $j$, $\vmax$, $\vpeak$, $a_{\rm vpeak}$, $a_{\rm first}$, $a_{\rm last}$, $n_{\rm sub}$, and $n_{\rm sat}$.
}
\label{fig:shuffle_inner}
\end{figure*}

The left-hand panel of Fig.~\ref{fig:shuffle_inner} shows the results of such a shuffling procedure applied to the $0.01 \hmpcc$ sample. The top panel shows the clustering of the original galaxy sample (black solid line) while the black dashed line is the clustering of the shuffled-by-mass sample.  The black line in the bottom panel plots the ratio of these two correlation functions,  representing the total level of GAB in this sample.  (We will discuss the additional measurement represented by the blue line in \S\ref{subsec:gab-inner} below.) The uncertainty on this measurement, estimated from 10 different shuffled samples (denoted by the shaded region), is negligible over most of the range and only noticeable at the largest separations.  On small scales, the clustering of the shuffled sample remains the same as the original one, since the contribution of galaxy pairs in the one-halo regime is unchanged.  The impact of GAB is clearly seen on large scales, where the clustering amplitude of the original shuffled is about 15 per cent above that of the shuffled sample.  As explained in \citet{Zehavi2018}, this increased clustering arises from the combined effect of halo assembly bias and the occupancy variations.  For example, for halo age, such a tendency arises from the preferential occupation of older haloes, at any fixed halo mass, which in turn exhibit stronger clustering.

In what follows, we set out to investigate the specific contributions to this GAB signal from the individual secondary properties.  Following \citet{Croton2007}, we extend the shuffling methodology in order to examine the role of different halo/environment properties. This is achieved by performing a doubly shuffled sample, where we randomly reassign the galaxies to haloes of the same mass and an additional specified halo property.  Shuffling in this manner removes all assembly bias effects other than those associated with this chosen property.  Comparing the clustering of this new shuffled sample to that of the mass-only shuffled sample and the clustering of the original sample will clearly indicate the importance of that parameter. For example, in the oversimplified case that GAB solely arises from one halo property, shuffling while holding that parameter and mass fixed will result in no difference to the clustering relative to the original one. Conversely, if this halo property has no bearing on GAB, the resulting clustering will be the same as the mass-only shuffling case.  Typically, the resulting clustering level will be somewhere between these two extremes, indicating the level of GAB imparted by this property.  Note that such an analysis does not take into account the correlations between different secondary properties. Rather, it informs in a clear and direct way the individual contribution of each secondary property to GAB.  

All the two-point correlation functions are calculated using \texttt{Corrfunc} \citep{Sinha2019,Sinha2020}, which is a fast tool to measure clustering statistics, using the programming language C with a Python interface.

\subsection{Internal halo properties}
\label{subsec:gab-inner}

The bulk of assembly bias studies have focused on studying the impact of halo properties such as formation time, concentration, spin, substructure, and others on the clustering of haloes (e.g. \citealt{Wechsler06,Gao07,Faltenbacher2010}). More recently, their impact on the halo occupation has also been explored \citep{Zehavi2018,Artale2018,Contreras2019,Bose2019}.  In this section, we study the ``percolation'' of these assembly bias effects to the clustering of galaxies. We explore the direct impact of the secondary halo properties on the GAB signature.

These secondary halo properties often correlate with halo mass, and their range of values varies by large amounts at different halo masses. Using bins of actual values of the secondary property will thus limit the sample that can be used for the shuffling. Instead, for each secondary property considered, we first rank the haloes by this property in narrow (0.1 dex) bins of halo mass. In each mass bin, we shuffle the galaxies according to their ranked secondary property in bins of 10 percentile. The result is a combined shuffling in bins of fixed mass and fixed (rank of) secondary property. Finally, we calculate the correlation function of the doubly-shuffled galaxy sample and compare it to the clustering of the original sample and that of the mass-only shuffled sample.  For clarity of presentation, we mostly plot the ratio of the correlation function of the doubly-shuffled sample to that of the mass-only shuffled one.

Returning to Fig.~\ref{fig:shuffle_inner}, the left-hand panel shows the clustering results for the galaxy sample shuffled by both mass and our measure of halo age $a_{\rm 0.5}$ (blue lines).  The top panel compares the correlation function for this sample to that of the original SAM galaxy sample and the mass-only shuffled one.  On the bottom panel, the black line is again the ratio of the original to mass-only shuffled clustering, manifesting the full GAB.  The blue line in the bottom panel denotes the ratio of the clustering of the sample shuffled by mass and $a_{\rm 0.5}$ to the clustering of the sample shuffled only by mass. This measure reflects the amount of GAB that can be attributed to this additional halo property. As discussed in \S3.1,  the closer this (blue) line is to the full GAB (black) line, the more of GAB is produced by this secondary property.  We quantify this using a simple measure of the fraction of the total GAB that can be attributed to a secondary property x:
\begin{equation}
  \label{eq:fab}
  f_{\rm AB} = \langle \, (\xi_{\rm sh(\Mh,x)}/\xi_{\rm sh(\Mh)} - 1) / (\xi_{\rm o}/\xi_{\rm sh(\Mh)} - 1) \, \rangle \,,
\end{equation}  
where the averaging is done over large scales (r $\sim 5-38 h^{-1}{\rm Mpc}$). Interestingly, for $a_{\rm 0.5}$  the resulting fraction is $f_{\rm AB}=0.26$, namely only 26 per cent of the full GAB can be accounted for by halo age.

The right-hand side of Fig.~\ref{fig:shuffle_inner} shows the clustering ratios for all the internal halo properties we consider, listed in \S2.2. We see that $a_{\rm 0.5}$, concentration, $\vmax$ and $\vpeak$, individually, each amount to only about 20 per cent of the total GAB, while the specific angular momentum has an even lower contribution. Among all the internal properties, the number of substructures and the (highly correlated to it) number of satellites appear to contribute the most at about 30 per cent. Again, we caution the reader that these properties are largely correlated with each other, and that these values reflect the individual fractions associated with each property, and should by no means be added up as independent values. It appears that none of these internal halo properties can account for the majority of the assembly bias, and as such none provide the full information needed for precise modelling of the large-scale clustering of galaxies.

\begin{figure}
	\centering
	\begin{subfigure}[h]{0.48\textwidth}
	    \includegraphics[width=\textwidth]{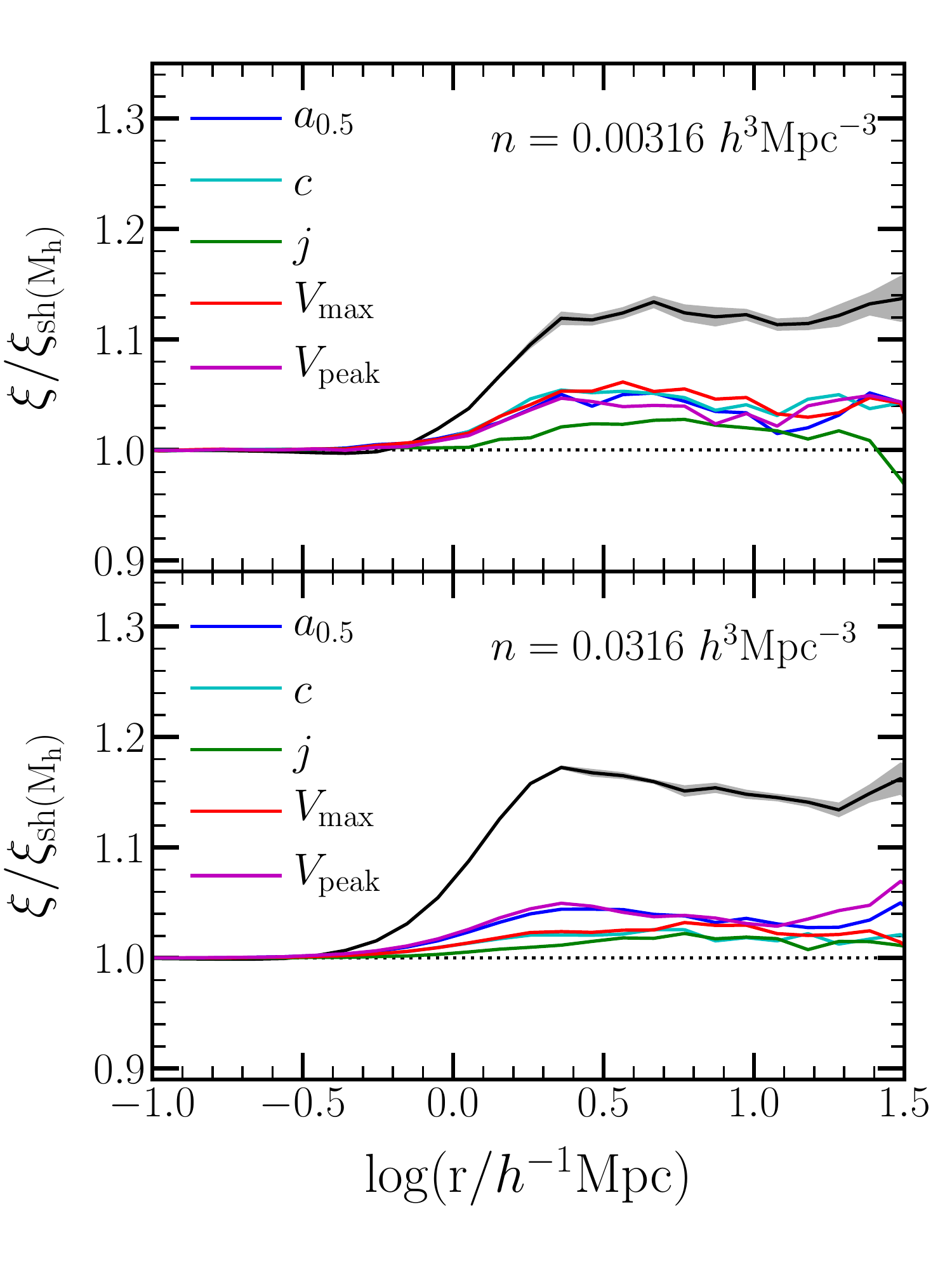}
    \end{subfigure}	    
	\hfill
\caption{
The same as the top right panel of Fig.~\ref{fig:shuffle_inner}, but for the number density samples corresponding to $n=0.00316 \hmpcc$ and $n=0.0316 \hmpcc$.
}
\label{fig:shuffle_inner_n13}
\end{figure}

Fig.~\ref{fig:shuffle_inner_n13} presents these clustering ratios for the two other number density samples we consider. We show here the five properties included in the top right panel of Fig.~\ref{fig:shuffle_inner}. The levels of GAB captured by these properties vary slightly with number density, generally decreasing with increased number density (lower stellar mass threshold).  However, the general trends and relative importance of these properties remain the same overall.  A similar picture is obtained for the other five parameters (not shown). Our conclusions thus appear robust to the number density of the sample. The $f_{\rm AB}$ values for all ten internal halo properties we consider and for the three number density samples are listed in the left-most part of Table~\ref{table:table1}. Again, these are calculated according to Eq.~\ref{eq:fab} and represent the fraction of GAB corresponding to each set property.  

\begin{table*}
  \caption{The fraction $f_{\rm AB}$ of the total GAB accounted for by the different secondary properties we consider, computed using Eq.~\ref{eq:fab}. All internal halo properties and environmental measures utilized are listed, and we quote the $f_{\rm AB}$ values for the three number densities samples, $n_1=0.00316 \hmpcc$, $n_2=0.01 \hmpcc$, and $n_3=0.0316 \hmpcc$.}
\centering
\setlength\extrarowheight{5pt}
\setlength{\tabcolsep}{6pt}
 \begin{tabular}{c c c c | c c c c |c c c c } 
 \hline
 halo prop & $f_{\rm AB}(n_1)$ & $f_{\rm AB}(n_2)$ & $f_{\rm AB}(n_3)$ & halo prop & $f_{\rm AB}(n_1)$ & $f_{\rm AB}(n_2)$ & $f_{\rm AB}(n_3)$ & halo prop & $f_{\rm AB}(n_1)$ & $f_{\rm AB}(n_2)$ & $f_{\rm AB}(n_3)$\\ 
 \hline\hline
 $a_{\rm 0.5}$ & 0.23 & 0.26 & 0.20 & $\delta_{\rm 1.25}$ & 0.74 & 0.88 & 0.98 & $\alpha_{\rm 0.3,1.25}$ & 0.90 & 1.02 & 1.03 \\ \hline
 $c$ & 0.28 & 0.21 & 0.09 & $\delta_{\rm 2.5}$ & 1.08 & 1.27 & 1.32 & $\alpha_{\rm 0.3,2.5}$ & 0.61 & 0.73 & 0.72 \\
 \hline
 $j$ & 0.05 & 0.05 & 0.06 & $\delta_{\rm 5}$ & 1.30 & 1.41 & 1.44 & $\alpha_{\rm 0.3,5}$ & 0.29 & 0.39 & 0.35  \\
 \hline
 $\vmax$ & 0.29 & 0.19 & 0.12 & $\delta_{\rm 10}$ & 1.20 & 1.32 & 1.31 & $\alpha_{\rm 0.3,10}$ & 0.04 & 0.05 & 0.09 \\
 \hline
 $\vpeak$ & 0.23 & 0.20 & 0.23 & $\alpha_{\rm 1, 1.25}$ & 0.13 & 0.06 & -0.06 &  $\alpha_{\rm 0.55,1.25}$ & 0.72 & 0.71 & 0.62  \\ 
 \hline
 $a_{\rm vpeak}$ & 0.08 & 0.07 & 0.04 & $\alpha_{\rm 1, 2.5}$ & -0.15 & -0.15 & -0.12 & $\alpha_{\rm 0.55,2.5}$ & 0.25 & 0.28 & 0.20 \\ 
 \hline
 $a_{\rm first}$ & -0.05 & 0.01 & 0.01 & $\alpha_{\rm 1, 5}$ & -0.04 & -0.02 & 0.02 & $\alpha_{\rm 0.55,5}$ & 0.04 & 0.02 & 0.01 \\
 \hline
 $a_{\rm last}$ & 0.09 & 0.05 & 0.05 & $\alpha_{\rm 1, 10}$ & 0.12 & 0.15 & 0.21 & $\alpha_{\rm 0.55,10}$ & 0.01 & -0.03 & -0.02 \\
 \hline
 $n_{\rm sub}$ & 0.39 & 0.32 & 0.31 & $\alpha_{\rm type(1.25)}$ & 0.03 & -0.01 & 0.12 & $r_{\rm 10}$ & 0.31 & 0.43 & 0.47 \\
 \hline
 $n_{\rm sat}$ & 0.34 & 0.29 & 0.24 & $\alpha_{\rm type(2.5)}$ & -0.04 & 0.09 & 0.21 \\
 \hline
 & & & & $\alpha_{\rm type(5)}$ & 0.33 & 0.47 & 0.56 \\ 
 \hline 
 & & & & $\alpha_{\rm type(10)}$ & 0.62  & 0.73 & 0.75 \\ [0.5ex]
 \hline
\end{tabular}
\label{table:table1}
\end{table*}

Our results are consistent with previous studies that indicated that single halo internal properties contribute only a small fraction of the measured GAB. Using a similar shuffling test, based on an older SAM applied to the Millennium simulation, \citet{Croton2007} demonstrated that neither formation redshift nor concentration encodes sufficient information to account for GAB.  Our findings verify their earlier result and extend it to a wide range of halo properties. In a recent work, \citet{Hadzhiyska2020a} use an abundance-matching inspired methodology to maximally reassign galaxies to haloes according to secondary properties using the IllustrisTNG hydrodynamical simulation \citep{Nelson2019}. Similarly, they find that properties like age, concentration, or spin impart only a small change to the level of clustering on large scales.

\subsection{Environmental properties}
\label{subsec:gab-env}

The environment of the haloes, often measured as the dark matter density field smoothed on a given scale, strongly impacts halo clustering. For example, in excursion set theory \citep{Bond91}, it is more likely to form a halo of specific mass in a denser background where it is easier to reach the collapse threshold, and thus a higher halo bias in those regions.
Besides the early works on assembly bias which focused on the dependence of halo clustering on internal halo properties, recent studies show that different environmental measures, such as the matter density field, cosmic web type, and tidal anisotropy have a major role in HAB (e.g. \citealt{Han2019, Ramakrishnan2019}). Measured galaxy clustering also strongly depends on these parameters, however most of this dependence can be explained using mass-only HOD models indicating the effect on galaxy properties to be small \citep{Abbas05,Abbas06,Paranjape2018b,Alam2019}. 
Moreover, when examining the occupancy variations, the large-scale environment introduces a much weaker dependence in the HOD relative to that of internal halo properties like halo formation time \citep{Mehta2014,McEwen2018,Zehavi2018,Artale2018}. Despite the weak environment OV, the halo clustering dependence on density (and tidal anisotropy) is strong, such that galaxy clustering may still be largely affected. Here we investigate the contribution of different measures of the environment to the full GAB signature, using the same double-shuffling methodology.

\begin{figure*}
	\centering
	\begin{subfigure}[h]{0.48\textwidth}
		\includegraphics[width=\textwidth]{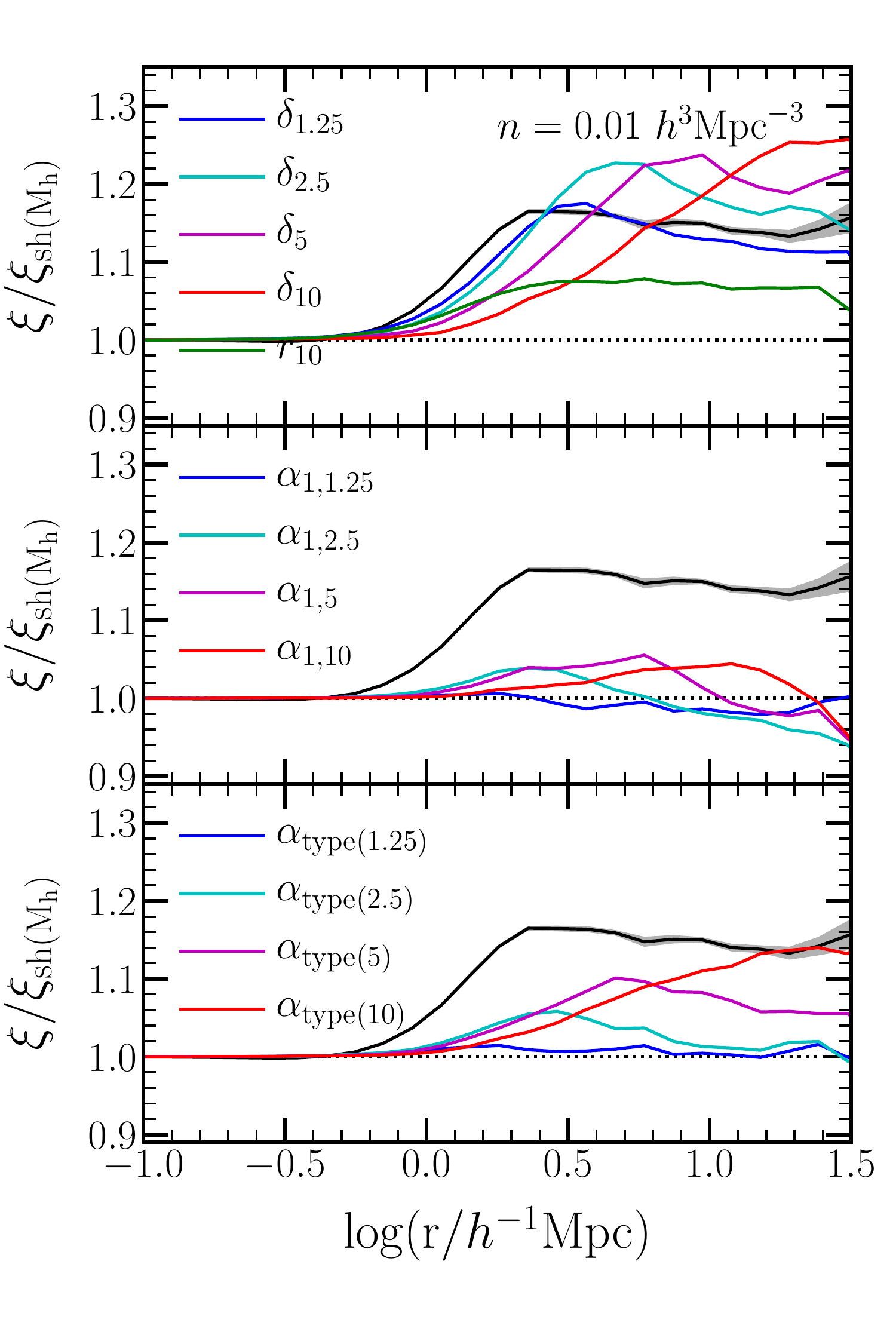}
	\end{subfigure}
	\hfill
	\begin{subfigure}[h]{0.48\textwidth}
		\includegraphics[width=\textwidth]{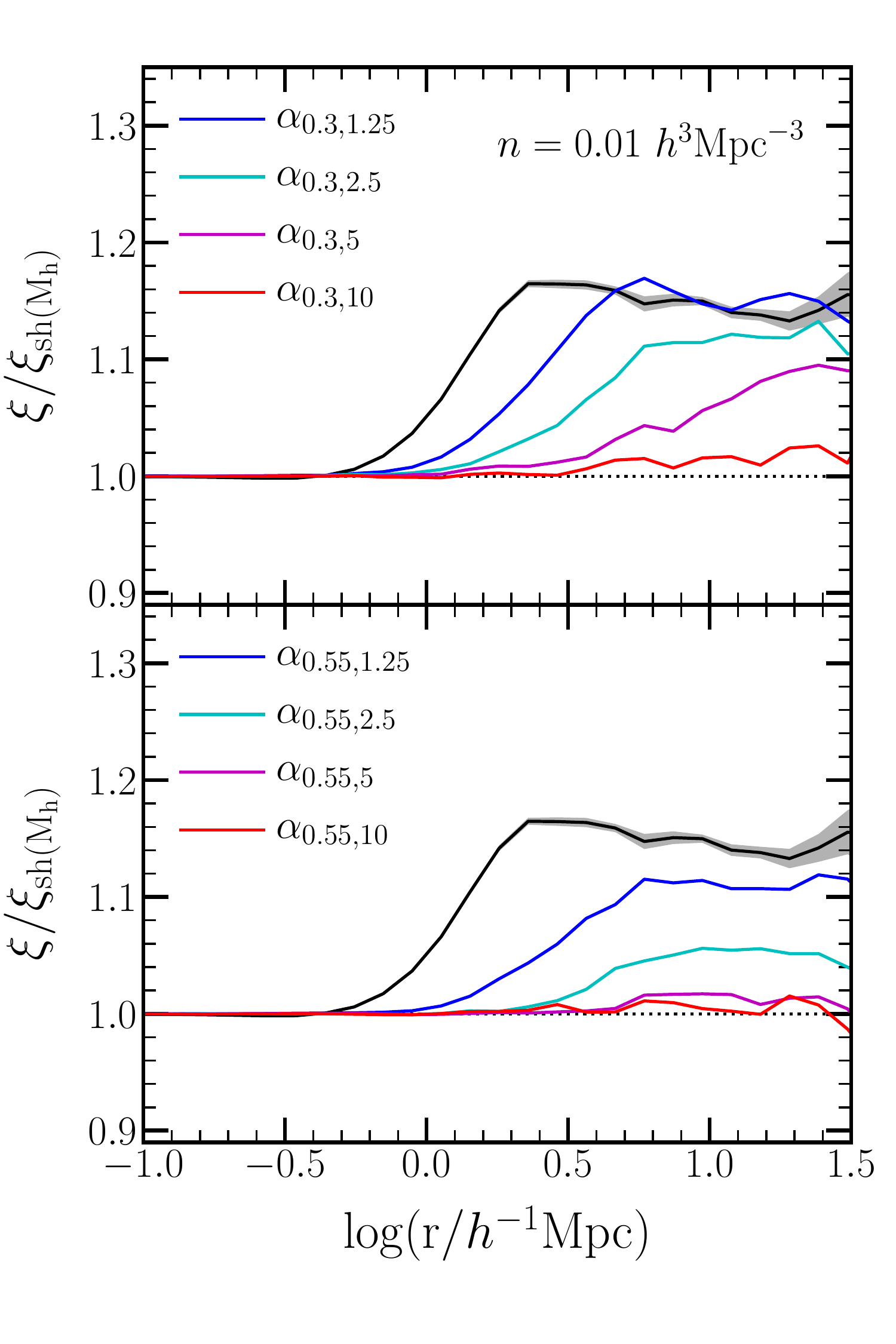}
	\end{subfigure}
	\hfill
\caption{Same as the right panel of Fig.~\ref{fig:shuffle_inner}, but now the second properties are the various environmental measures, including: $\delta_{\rm 1.25/2.5/5/10}$ and $r_{\rm 10}$ (top left), $\alpha_{\rm 1, 1.25/2.5/5/10}$ (middle left), $\alpha_{\rm type(1.25/2.5/5/10)}$ (bottom left), $\alpha_{\rm 0.3, 1.25/2.5/5/10}$ (top right) and $\alpha_{\rm 0.55, 1.25/2.5/5/10}$ (bottom right).
}
\label{fig:xienv}
\end{figure*}

Fig.~\ref{fig:xienv} shows the clustering measurements when shuffling by both halo mass and an additional environment parameter, using a variety of different estimators. The results are shown relative to the clustering of the mass-only shuffled sample, in a similar manner to that in Fig.~\ref{fig:shuffle_inner}. Again, we first focus on the stellar-mass threshold sample corresponding to a number density of $0.01 \hmpcc$. The top panel on the left-hand side of Fig.~\ref{fig:xienv} presents the results for smoothed density fields obtained from the distribution of dark matter particles in the simulation using different Gaussian smoothing scales, ranging from 1.25 to 10 $\hmpc$ ($\delta_{\rm 1.25/2.5/5/10}$).

The coloured solid curves show the GAB recovered by doubly-shuffled samples at fixed mass and the different smoothed densities, and the black solid line with the shaded region again shows the full GAB level in the original SAM sample for comparison. All density measures considered recover the full level of GAB or higher on large scales, and the GAB level increases with smoothing scale. This is in stark contrast to the internal halo properties case.  Among the smoothing scales, $\delta_{\rm 1.25}$ has the closest GAB level to that of the SAM on the relevant scales (with $f_{\rm AB} \sim 0.9$ for this sample). In other words, shuffling while holding $\delta_{\rm 1.25}$ and halo mass fixed results in close to the same level of clustering as that of the original sample. Note that this does not necessarily imply that $\delta_{\rm 1.25}$ is the sole dominant factor in characterizing GAB. In fact, we will show that another definition of the environment, such as $\alpha_{\rm 0.3,1.25}$ (see below), can also reproduce the full GAB signal. 

One must be careful when interpreting these findings. In some sense, the relation of environment to GAB may seem obvious, since by its nature stronger clustering corresponds to denser environments.  In detail, however, this correspondence is not trivial since we are exploring here the contribution of secondary properties on top of the primary halo mass environmental dependence. Moreover, it is not a-priori known which aspect or measure of the environment can best capture this dependence. Furthermore, the halo environment may relate more directly to halo clustering, namely HAB, while we are concerned here with the overall effect on galaxy clustering, arising from the combined effect of HAB and the occupancy variations. In any case, our aim here is not necessarily to explain the origin of GAB, but rather to find the most informative halo property to characterize it.  This will allow to incorporate GAB into the HOD formalism and utilize it to produce realistic mock catalogues that include assembly bias, as we discuss in detail in \S~\ref{sec:edhod}.

Regarding the increased GAB signal obtained for the densities with larger smoothing scales, we see that holding these properties fixed results in stronger clustering. This implies that in the original sample the galaxies are not assigned to haloes according to $\delta_{\rm 2.5/5/10}$ in a tight relation. So when reassigning galaxies while fixing halo mass and $\delta_{\rm 2.5/5/10}$ the clustering increases. Taking another point of view, shuffling at fixed $\delta_{\rm 1.25}$ perhaps reproduces the original clustering since it contains information from both the larger scales (outside a halo) and the internal smaller scales, while shuffling at fixed density on larger scales may ignore information from some internal or environmental properties that can suppress the clustering. The results for the large smoothing scales also seem directly related to the somewhat unnatural ``carving out'' of regions of that size.  This is seen clearly in the top panel of fig.~1 of \citet{Zehavi2018}, which shows the distribution of haloes in a slice from the simulation colour coded by $\delta_{\rm 5}$.  This is also the reason that the scale at which the maximum GAB is obtained corresponds directly to the smoothing scale. The clustering strength on scales larger than the smoothing scale are also boosted, since the overdense (or underdense) regions of the specific smoothing scale are not fully independent but rather correlated with the densities measured on larger scales.

Before proceeding to investigate the tidal anisotropy, we first examine a simply defined environment property $r_{\rm 10}$, which is a measure of the tidal force from a nearby massive neighbour that impacts the halo in consideration the most (see \S~\ref{subsec:properties}).  This measure is shown as the green solid line in the top-left panel of Fig.~\ref{fig:xienv}. We see that it does not work as well as the smoothed density measures, but in fact, with $f_{\rm AB}=0.43$, it still accounts for more of the GAB signal than the internal properties previously shown.  Table~\ref{table:table1} lists the fractional values for all the environmental parameters as well, in the middle and right sides of the Table.

\begin{figure}
	\centering
	\begin{subfigure}[h]{0.48\textwidth}
	    \includegraphics[width=\textwidth]{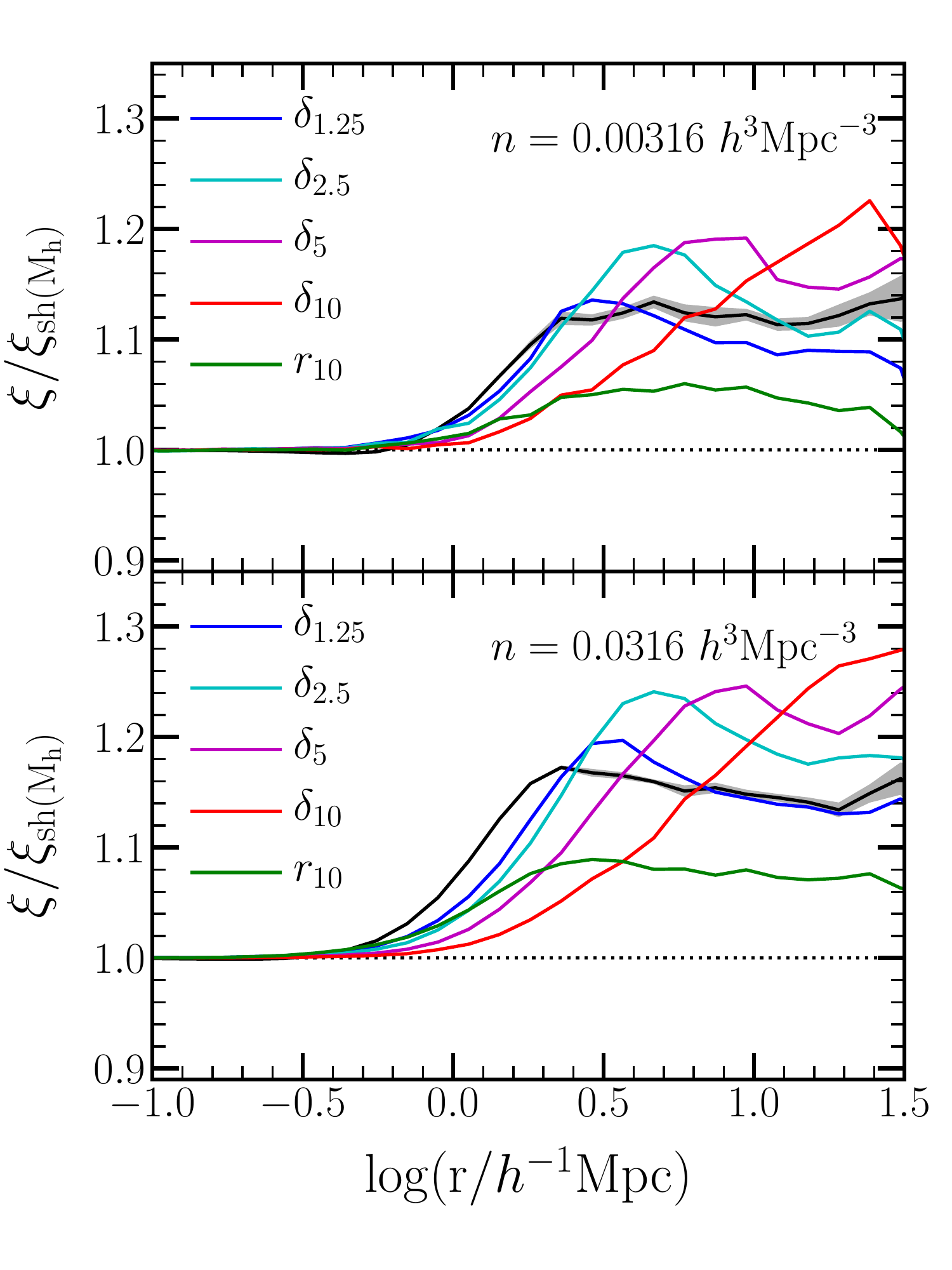}
    \end{subfigure}	    
	\hfill
\caption{
Same as the top-left panel of Fig.~\ref{fig:xienv}, but for the two other number density samples, $n=0.00316 \hmpcc$ and $n=0.0316 \hmpcc$.
}
\label{fig:xienv_n13}
\end{figure}

The results shown thus far regarding the importance of the environmental properties are based on the galaxy sample with number density of $n_2=0.01 \hmpcc$. We repeat the same analysis for the other two number density samples and report the results in Fig.~\ref{fig:xienv_n13} and Table~\ref{table:table1}. Fig.~\ref{fig:xienv_n13} shows the GAB level corresponding to $\delta_{\rm 1.25/2.5/5/10}$ and $r_{10}$ (the same as those in the top left panel of Fig.~\ref{fig:xienv}) for the two other density thresholds, $n_1=0.00316 \hmpcc$ and $n_3=0.0316 \hmpcc$. The general behaviours are the same for all number densities.  Interestingly, the GAB fractions for the environmental properties tend to increase with number density, opposite to the trend for the internal properties. Most strikingly, for all samples, the contributions to GAB from the densities are much higher than that of the halo internal properties.

There is increasing evidence that the tidal anisotropy is a key factor in determining HAB \citep{Paranjape2018a,Ramakrishnan2019,Mansfield2020}. Here we would like to explore whether the tidal anisotropy and related measures like the cosmic web type are also determining factors for GAB. Our definition of the tidal anisotropy parameter $\alpha$ (Eq.~\ref{eq:alpha}) allows to vary the normalization power, however we start with $\alpha_1$, following \citet{Paranjape2018a} and \citet{Ramakrishnan2019}.

The middle panel of the left-hand side of Fig.~\ref{fig:xienv} shows the GAB level recovered when shuffling by both halo mass and this tidal anisotropy parameter, for our four smoothing scales. 
We find that this parameter contributes a very small (or even negative) fraction to the GAB of the original sample. Namely, the tidal anisotropy parameter defined in this manner has very little impact on the clustering and in a couple of the cases even tends to reduce it slightly. We caution that our smallest smoothing scale (1.25 $\hmpc$ Gaussian, roughly corresponding to a $\sim\sqrt{5}\times1.25$ top-hat), is different than the adaptive $4R_{\rm 200}$ smoothing scale used by \citet{Paranjape2018a}. We find that HAB shows different dependences on $\alpha_{1}$ measured for different fixed scales (see Appendix~\ref{sec:appendix}), and as such an adaptive scale of $R=4R_{\rm 200}$ may still be important for GAB (R.\ Sheth and A.\ Paranjape, private communication). We explore alternate definitions of the tidal anisotropy parameter shortly below.

In the bottom panel of the left column of Fig.~\ref{fig:xienv}, we consider the GAB levels associated with cosmic web type measured from the sign of the eigenvalues of the tidal tensor, classifying the large-scale structure to different environments. Instead of shuffling at fixed mass and fixed rank of the secondary property, we shuffle galaxies in haloes of the same mass and within the same broad category of nodes, filaments, sheets, and voids.  Again, we use the four different smoothing scales to calculate the tidal tensor and obtain $\alpha_{\rm type (1.25/2.5/5/10)}$. We find that the GAB level increases with the smoothing scale, such that while $\alpha_{\rm type(1.25)}$ is essentially unrelated to it, $\alpha_{\rm type(10)}$ corresponds to about 73 per cent of the GAB signature on large scales. This $f_{\rm AB}$ value translates to a ratio of 0.96 between the correlation function at fixed mass and cosmic web type and the original correlation function of the SAM galaxy sample, in general agreement with a similar analysis performed by \citet{Hadzhiyska2020a} using IllustrisTNG. Remaining differences may be attributed to their modified cosmic web type definition and the details of the smoothing applied. Note that while $\alpha_{\rm type(10)}$ matches the SAM GAB on the largest scales, it significantly deviates from it on intermediate scales, so it is not an ideal environmental property for characterising GAB in any case. 

Motivated by the relative importance of the cosmic-web type environment, we set to explore whether a modification of the normalisation power $n$ of the tidal anisotropy parameter (Eq.~\ref{eq:alpha}) may prove more useful. Other works \citep{Paranjape2018a,Alam2019} experimented with this aiming to minimize the correlation with $\delta_R$. Here we try different values of $n$, but instead aim to best reproduce GAB. In the right-hand side of Fig.~\ref{fig:xienv}, we present the GAB level associated with a couple different such values.  The top panel shows our most successful case with a normalisation power of 0.3, still plotting $\alpha_{\rm 0.3, 1.25/2.5/5/10}$ for our four different smoothing scales, and the bottom panels shows the results for $n=0.55$, the value adopted by \citet{Alam2019}. In both cases, the GAB level decreases with increasing smoothing scale, opposite to the cosmic-web case and also that of $\delta$. For the normalisation power of $n=0.55$, the highest level of GAB is obtained for $\alpha_{\rm 0.55,1.25}$, recovering $\sim$70 per cent of the total signal, while for $n=0.3$, $\alpha_{\rm 0.3, 1.25}$ reproduces roughly the full level of GAB.

It appears that the tidal anisotropy can potentially be utilized for describing GAB, though care must be given to the exact definition and smoothing scale used. We note that, as defined, $\alpha_{\rm 0.3, 1.25}$ does depend on the mass density $\delta_{\rm 1.25}$, however it clearly includes additional aspects of the environment captured by the tidal shear. 
Table~\ref{table:table1} includes as well the $f_{\rm AB}$ values for all the tidal properties, also for the two other number densities (not shown). The relative fractions of the GAB associated with the different properties depend somewhat on the stellar mass threshold (or number density) of the samples, but in general $\delta_{\rm 1.25}$ and $\alpha_{\rm 0.3,1.25}$ remain the most important properties that can recover the full level of GAB. In Section~\ref{sec:edhod} we will use both properties to model GAB.  

\section{Occupancy variations, HOD parameters and mock catalogues}
\label{sec:hodmock}

Having determined that both the density measured with $1.25 \hmpc$ Gaussian smoothing, $\delta_{\rm 1.25}$, and the tidal anisotropy parameter, $\alpha_{\rm 0.3,1.25}$, can best recover the full level of GAB,  we proceed to examine their occupancy variations and the related secondary trends in the stellar mass-halo mass relation. We then obtain HOD parametric fits, and demonstrate their usefulness in producing mock catalogues that incorporate the correct level of assembly bias.

\subsection{Occupancy variation with environment}
\label{subsec:hod}

As previously mentioned, GAB is the combined effect of the halo clustering dependence on secondary variables (namely HAB) and the variations of the galaxy occupation of haloes with these parameters (namely, the occupancy variations, or OV). In order to gain a better physical insight on using $\delta_{\rm 1.25}$ or $\alpha_{\rm 0.3,1.25}$ to describe GAB,  we start by exploring their halo occupation and OV.

\begin{figure*}
	\centering
	\begin{subfigure}[h]{0.48\textwidth}
		\includegraphics[width=\textwidth]{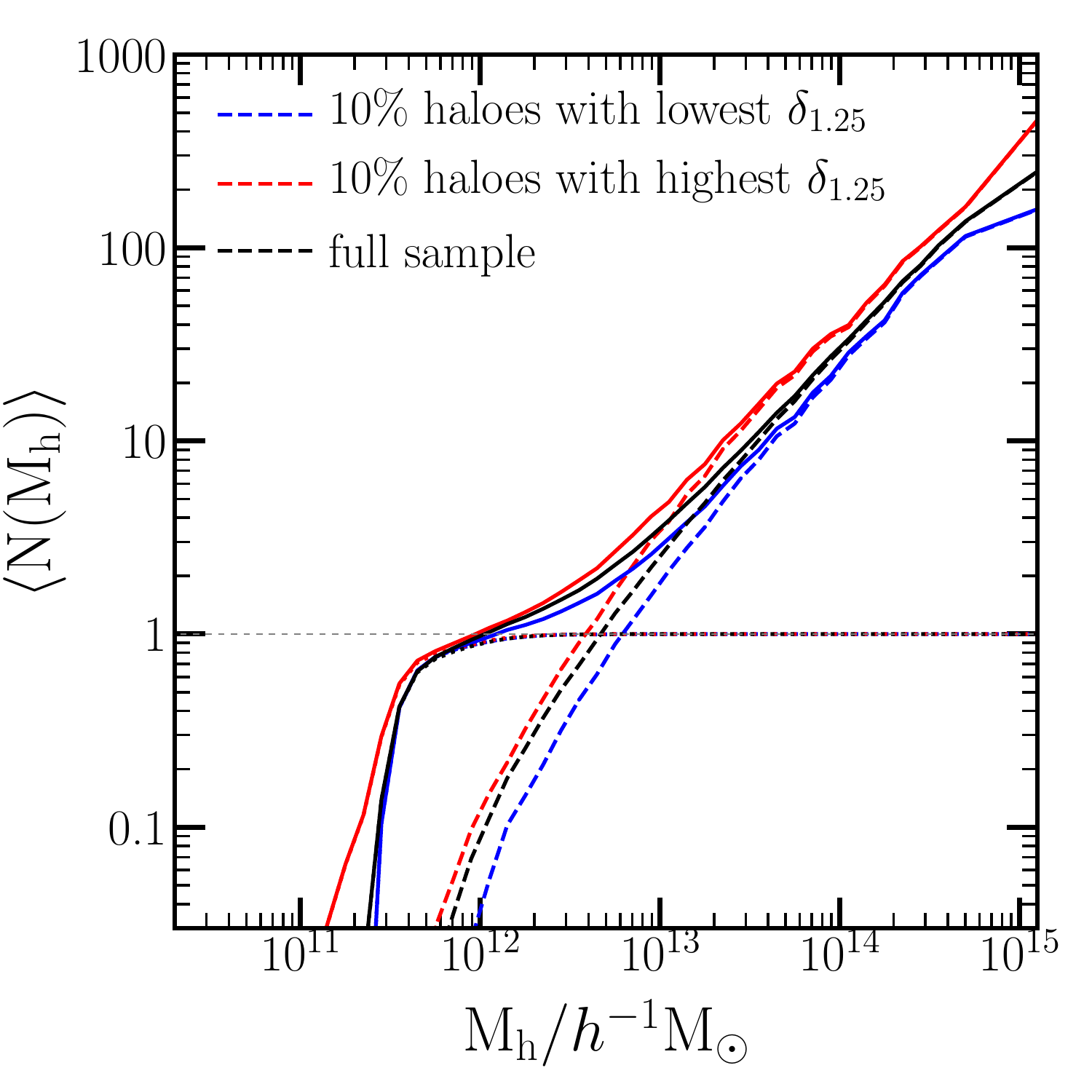}
	\end{subfigure}
	\hfill
	\begin{subfigure}[h]{0.48\textwidth}
        \includegraphics[width=\textwidth]{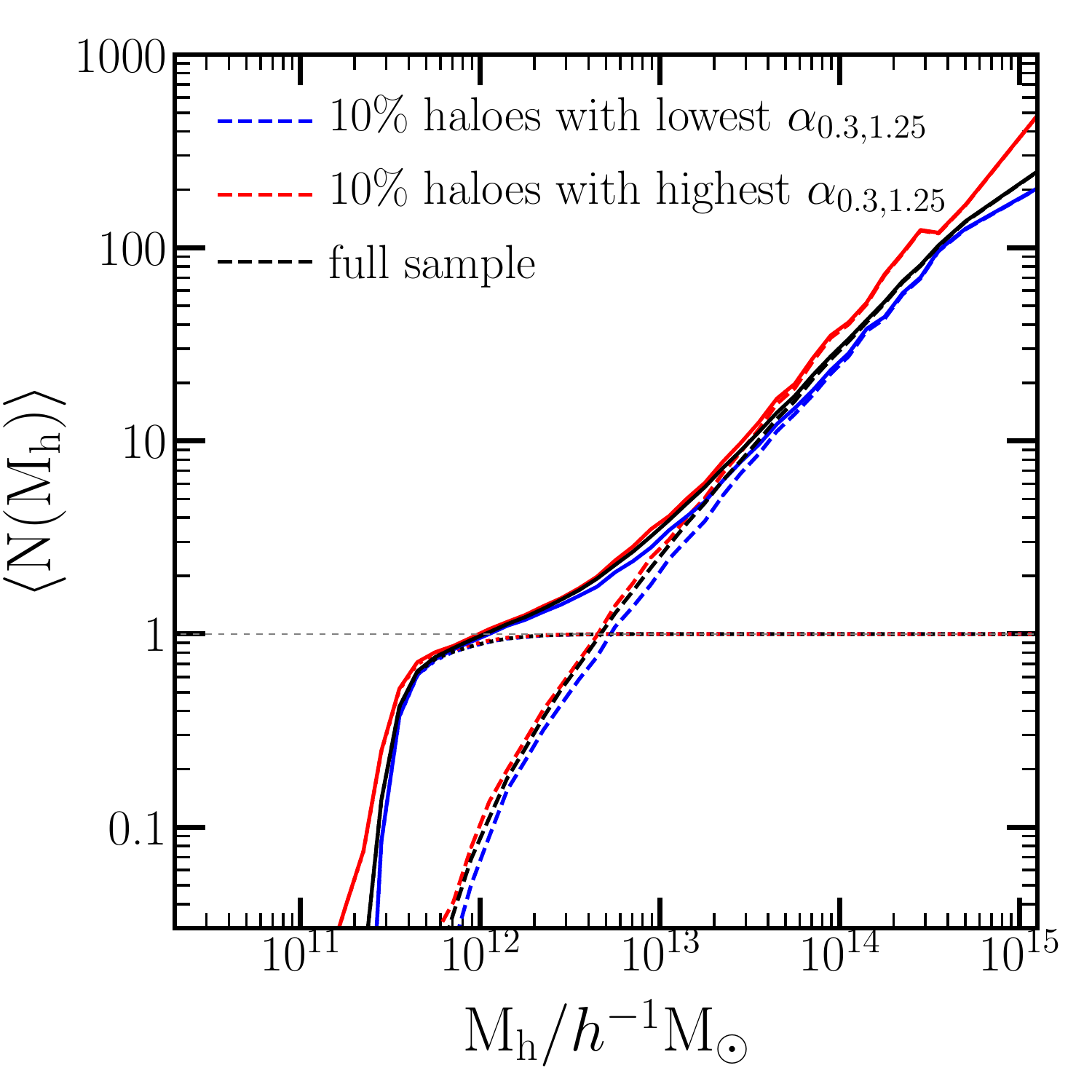}
	\end{subfigure}
	\hfill
\caption{Occupancy variations for $\delta_{\rm 1.25}$ (left) and $\alpha_{\rm 0.3,1.25}$ (right) for the $n=0.01 \hmpcc$ sample. The average number of galaxies as a function of halo mass is shown as the solid lines. The central galaxies and satellites occupations are shown as dotted and dashed lines, respectively. The occupation function for the full galaxy sample is shown in black, for galaxies in the 10 per cent of haloes with the highest values of the environmental property in red, and for galaxies in the 10 per cent of haloes with the lowest values in blue.
}
\label{fig:HOD125}
\end{figure*}

The fundamental statistic we utilize here is the halo occupation function, i.e the average number of galaxies as a function of halo mass. The solid black curves in Fig.~\ref{fig:HOD125} (the same in both panels) show this function for the $n=0.01 \hmpcc$ galaxy sample, measured directly from the SAM galaxy and halo catalogues. The dotted and dashed black curves separately show the contribution to the halo occupation from central galaxies and satellites, respectively. The occupation of central galaxies is similar to a smoothed step function, in which haloes gradually transition to hosting a central galaxy above a certain halo mass. The satellite occupation roughly follows a power low above a certain halo mass threshold. 

The standard HOD considers only the dependence of the halo occupation on halo mass. The dependence of this occupation function on secondary properties opens the way to assembly bias effects on galaxy clustering. \citet{Zehavi2018} discussed in detail the connection of OV to assembly bias and examined the OV with halo age and environment. They showed that at fixed halo mass, early-formed haloes are more likely to host a central galaxy and have fewer satellites galaxies above a stellar-mass threshold. \citealt{Contreras2019} extended this to higher redshifts and examines as well the OV with halo concentration. \citet{Zehavi2018} also investigated the environmental effects on the HOD using $\delta_{\rm 5}$ with the same simulation and SAM used here, finding subtle but distinct trends, such that haloes in dense environments start hosting central galaxies at lower halo mass, and have more satellites at fixed halo mass. 

We extend their study by considering different smoothing scales of the density field, as well as the tidal anisotropy variables discussed in \S~\ref{subsec:gab-env}. To the best of our knowledge, this is the first study of the OV with tidal anisotropy. Fig.~\ref{fig:HOD125} shows the dependence of the HOD on $\delta_{\rm 1.25}$ (left) $\alpha_{\rm 0.3,1.25}$ (right) for the $n=0.01 \hmpcc$ sample. As before, we rank the haloes according to the environmental property in fixed fine bins of halo mass. The red curves denote the occupation functions for galaxies in the 10 per cent of haloes with the highest environment measure, while the blue curves are the occupation function for galaxies in the 10 per cent of haloes with the lowest environment parameter.  Dotted and dashed curves again correspond to central galaxies and satellites, respectively. 

We find similar, but slightly stronger, trends of OV for $\delta_{\rm 1.25}$ relative to that shown in \citet{Zehavi2018} for $\delta_{\rm 5}$, as to be expected for the smaller smoothing length. The right-hand side of Fig.~\ref{fig:HOD125} shows the OV for $\alpha_{\rm 0.3,1.25}$.  We find that haloes with higher $\alpha_{\rm 0.3,1.25}$ (more anisotropic regions) begin to host central galaxies at lower halo mass, and have more satellite at a fixed halo mass. More anisotropic regions correspond to nodes or filaments, while low anisotropy regions correspond to voids (see Appendix~\ref{sec:appendix} for more details). It is expected that haloes in strong tidal anisotropy regions tend to host more galaxies, since the density is also higher in those regions. The amplitude of the $\alpha_{\rm 0.3,1.25}$ OV is, however, weaker than that of $\delta_{\rm 1.25}$. 

In Appendix~\ref{sec:appendix}, we present the OV with $\delta$ and $\alpha_{0.3}$ for the four smoothing lengths available to us, and confirm that the level of deviations decreases with increasing smoothing scale for both parameters. We also find, for any given smoothing scale, weaker OV with $\alpha_{0.3}$ than with $\delta$. For the two largest smoothing scales, in fact, the tidal anisotropy OV is nearly negligible.
For completeness, we also investigate in Appendix~\ref{sec:appendix} the OV dependence of $\alpha_{\rm 1}$ for the different smoothing scales. Interestingly, we find an opposite trend relative to that of $\delta$ and $\alpha_{\rm 0.3,1.25}$,  such that there is a preference for hosting centrals in low (more isotropic) $\alpha_1$ regions. This may arise from a reverse correlation with $\delta$, namely that high $\alpha_{\rm 1}$ corresponds to underdense regions,  which is reflected as well in a reverse HAB trend seen for the large smoothing scales. Taken together, these trends explain the GAB results shown in the middle-left-hand panel of Fig.~\ref{fig:xienv}.

\begin{figure*}
	\centering
	\begin{subfigure}[h]{0.48\textwidth}
		\includegraphics[width=\textwidth]{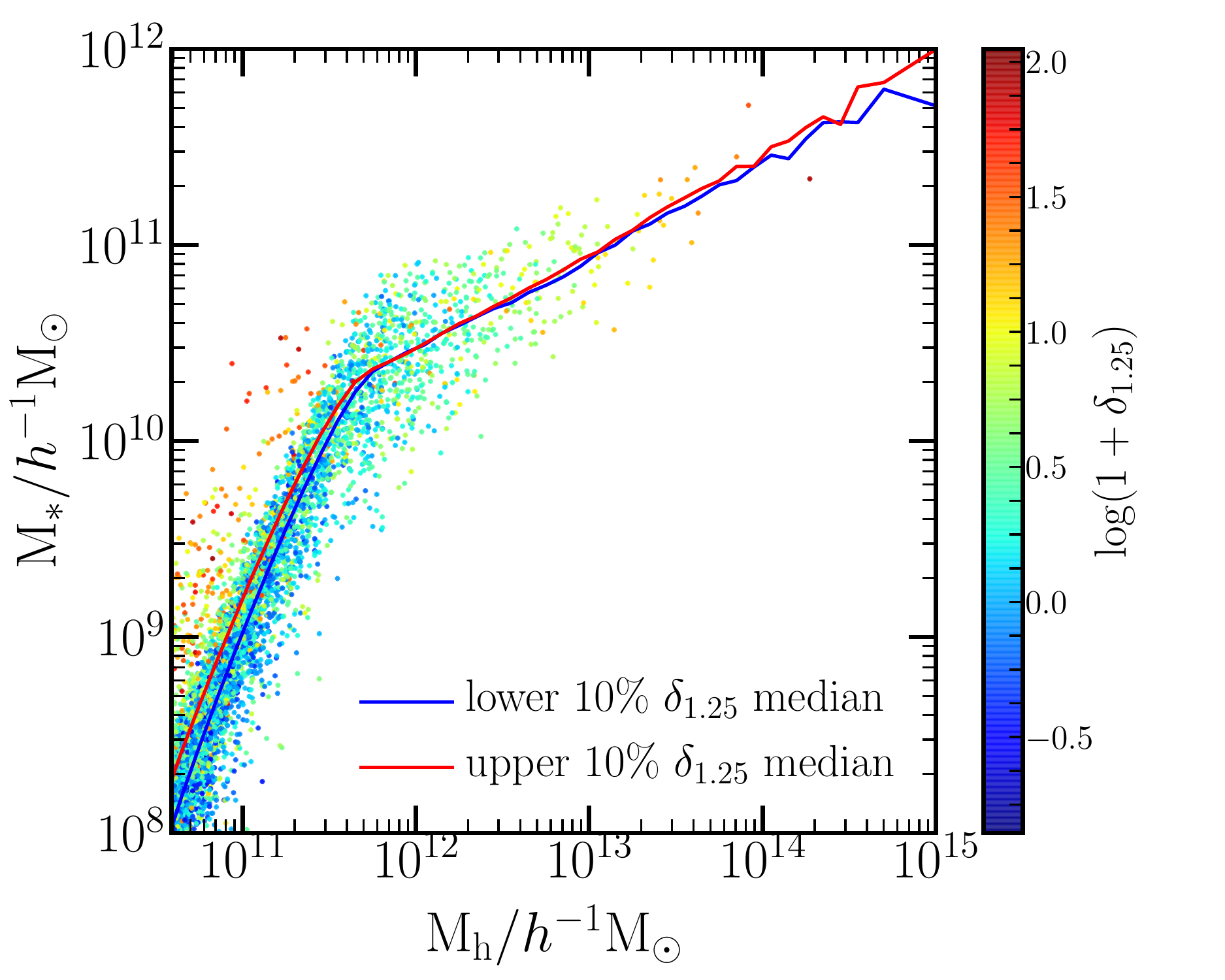}
	\end{subfigure}
	\hfill
	\begin{subfigure}[h]{0.48\textwidth}
		\includegraphics[width=\textwidth]{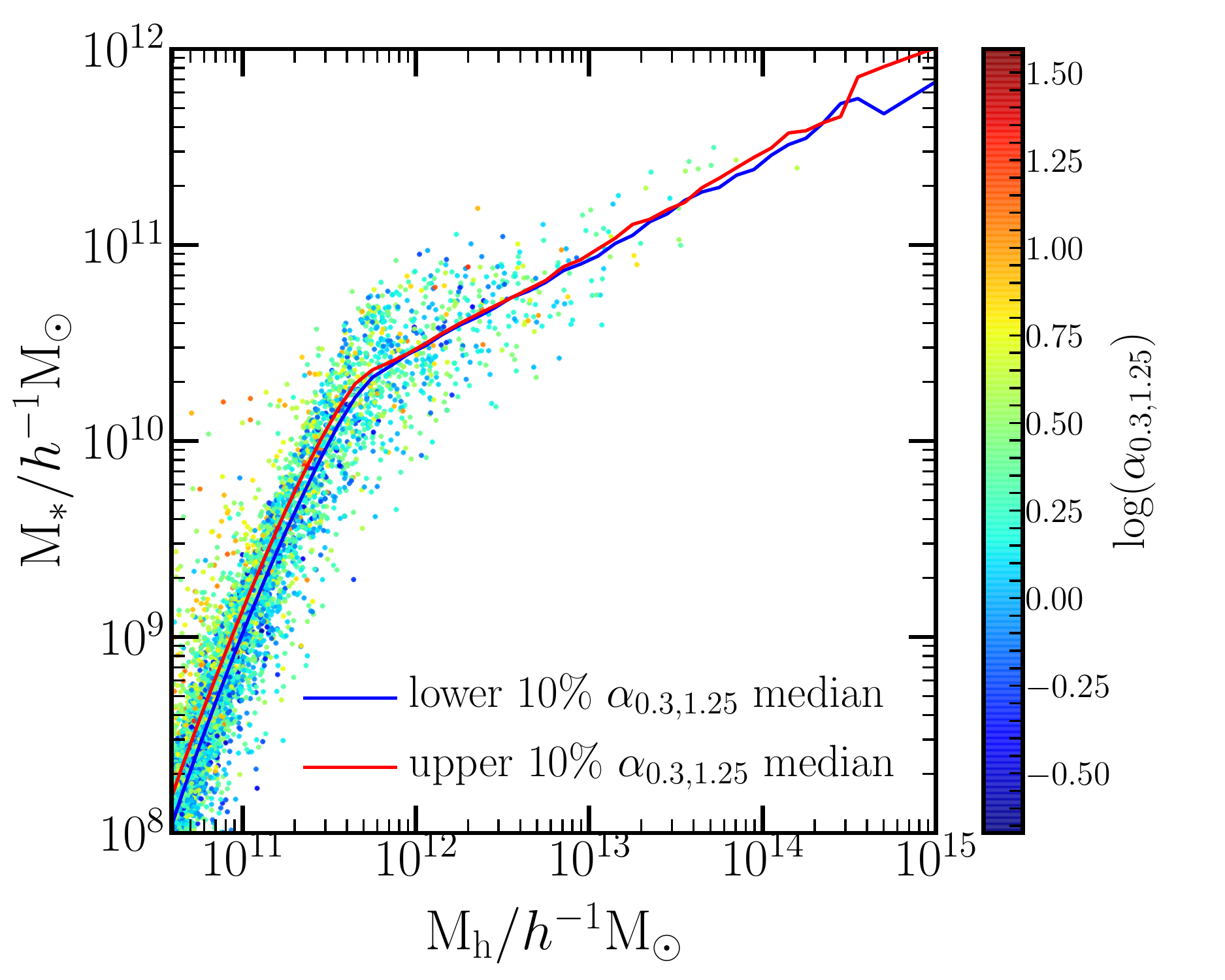}
	\end{subfigure}
	\hfill
\caption{The stellar mass-halo mass relation for central galaxies in the $n=0.01 \hmpcc$ sample and its environmental dependence.  For clarity we plot a representative (randomly chosen) 0.1 per cent of the galaxies. Galaxies are colour-coded by the on $\delta_{\rm 1.25}$ (left) and $\alpha_{\rm 0.3,1.25}$ (right), colour coded by the value of $\delta_{\rm 1.25}$ (left) and $\alpha_{\rm 0.3,1.25}$ (right). The solid red (blue) lines represents the median value of $M_{*}$, in each mass bin, for the 10 per cent of the haloes that have the highest (lowest) $\delta_{\rm 1.25}$ or $\alpha_{\rm 0.3,1.25}$.
}
\label{fig:SMHM125}
\end{figure*}

The origin of the central galaxies OV can be understood by examining the secondary trends in the stellar mass-halo mass relation for the central galaxies \citep{Zehavi2018}. In Fig.~\ref{fig:SMHM125}, we show the stellar mass-halo mass relation for central galaxies colour-coded by $\delta_{\rm 1.25}$ (left) or $\alpha_{\rm 0.3, 1.25}$ (right), for a randomly chosen sample which contains 0.1 per cent of the SAM sample. The central galaxies' stellar mass increases as a function of halo mass, with a steeper relation below $5\times10^{11}\hmsun$ and a shallower slope above that. At fixed halo mass, the scatter in the stellar mass is not random but rather correlated with the secondary halo or environment properties in such a way that early formed haloes or haloes in denser regions tend to host more massive central \citep{Zehavi2018,Xu2020}.

The trends for environmental properties are much weaker (or saying it another way, there is more scatter in these secondary trends for environmental properties) than those for the halo internal properties. To see this trend more clearly, we show in red (blue) solid line the median value of $\Mstar$ of the haloes with the highest (lowest) 10 per cent environmental property at fixed halo mass. From these we indeed see that haloes in the highest density or highest tidal anisotropy regions tend to host slightly more massive centrals than those in the most underdense or lowest tidal anisotropy regions, at fixed halo mass. This implies that for any stellar-mass threshold used to define a galaxy sample (e.g. $1.42\times10^{10} \hmsun$ for our fiducial sample) galaxies preferentially occupy the haloes in the denser regions, producing the occupancy variations. The trend is the same for $\alpha_{\rm 0.3, 1.25}$ but with an even smaller difference between upper and lower 10 per cent lines, consistent with the smaller centrals OV for it.

Overall, the occupancy variations and the secondary trends in the stellar mass-halo mass relation for the environmental properties are smaller compared to that of halo age and concentration, but it is in fact this small trend coupled with the large HAB that account for essentially all of the GAB. To further verify this, we proceed to build mock catalogues based on $\delta$, $\alpha_{\rm 0.3}$, and concentration later in this section and compare the GAB level in them with that of the original sample.

\subsection{HOD parameters}
\label{subsec:hodparam}

\begin{figure*}
    \includegraphics[width=1.0\textwidth]{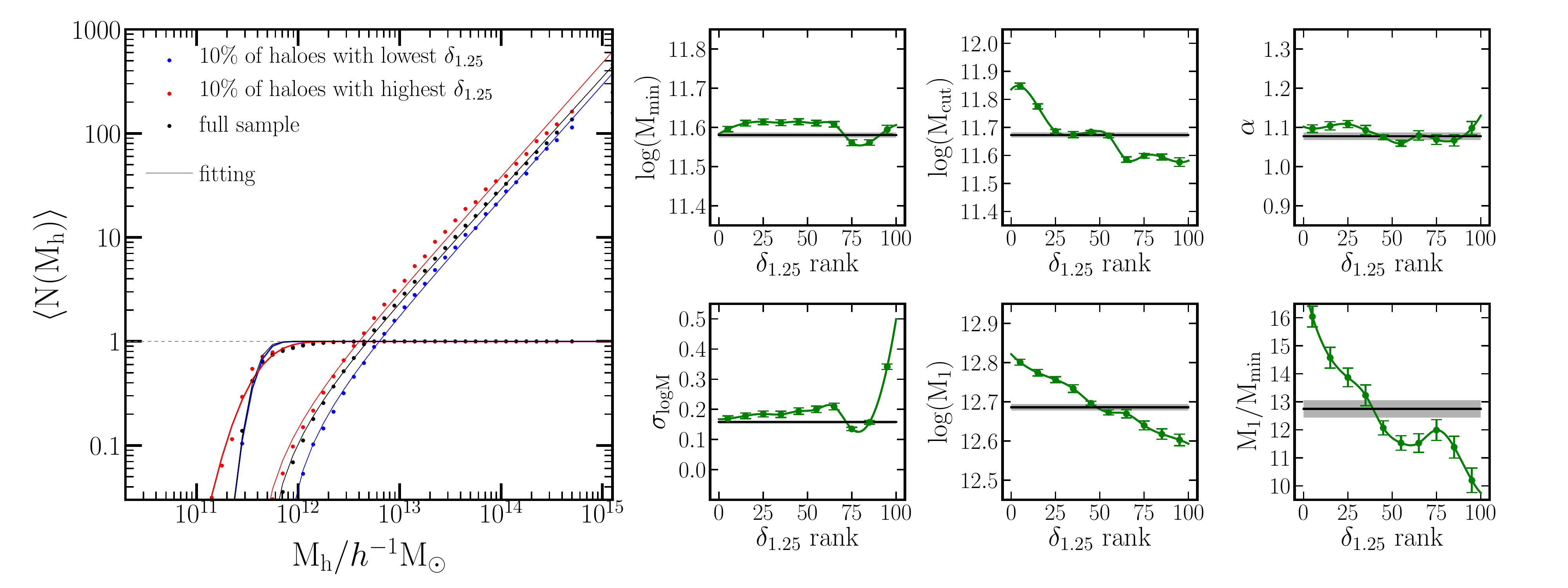}
\caption{Fitted HOD parameters as function of $\delta_{\rm 1.25}$. The panel on the left shows the HODs for different subsets of the $n=0.01 \hmpcc$ galaxy sample. The dots are the measured halo occupations shown separately for centrals and satellites, while the curves show the best-fits of the 5-parameter HOD model. Red, blue, and black represent the 10 per cent of haloes with the highest $\delta_{\rm 1.25}$, the lowest $\delta_{\rm 1.25}$, and the full sample, respectively. The panels on the right-hand side are the fitting results of the 5 individual parameters, shown as a function of the environment rank from 0 to 100. I.e, the green dots with error bars are the parameters inferred for each ranked 10 per cent subset of $\delta_{\rm 1.25}$. The green lines are spline fits between the parameter values, and the black lines with the shaded area are the ones inferred from the full sample. In addition to the 5 parameters, we show (as a 6th panel) the commonly used $M_{\rm 1}/M_{\rm min}$ parameter.
}
\label{fig:fithod_g125}
\end{figure*}

To better characterize the OV of environmental properties, it is also helpful to adopt a specific parametrization of the shape of the halo occupation function and investigate the dependence of the HOD parameters on the environmental properties.  For simplicity, we choose to focus here on the dependence on $\delta_{\rm 1.25}$, and we relegate $\alpha_{\rm 0.3, 1.25}$ to Appendix~\ref{sec:appendix2}.

For this purpose, we adopt the commonly used 5-parameter HOD model which is motivated by galaxy formation physics \citep{Zheng2005}. In this HOD model, the average number of central and satellites in a halo of mass $M_{\rm h}$ can be written as:
\begin{equation}
\label{eq:cen}
\langle N_{\rm cen}(\Mh)\rangle=\frac{1}{2} \left[ 1+{\rm erf} \left( \frac{{\rm log}\Mh+{\rm log}M_{\rm min}}{\sigma_{{\rm log}M}} \right) \right]
\end{equation}
and 
\begin{equation}
\label{eq:sat}
\langle N_{\rm sat}(\Mh)\rangle=\left( \frac{\Mh-M_{\rm cut}}{M^*} \right) ^\alpha.
\end{equation}
$M_{\rm min}$ is the characteristic halo mass for hosting a central galaxy, specifically set here as the halo mass for which on average half of the haloes are occupied. ${\sigma_{{\rm log}M}}$ is a scale parameter indicating the width of the transition in the central occupation, and reflects the scatter between stellar mass and halo mass. $M_{\rm cut}$ is the halo mass threshold above which the halo can host satellites, and $M^*$ measures the difference in halo mass that increases the number of satellites from 0 to 1. Another satellite parameter often used in literature is $M_1=M_{\rm cut}+M^*$, which characterizes the mass of haloes that host one satellite galaxy on average. The final HOD parameter is $\alpha$ (not to be confused with the tidal anisotropy $\alpha_{\rm n,R}$), the slope of the power-law for the satellites occupation. The total occupation function is then specified by these five parameters and expressed as the sum of these two terms:
\begin{equation}
\label{eq:censat}
\langle N(\Mh) \rangle = \langle N_{\rm cen}(\Mh) \rangle + \langle N_{\rm sat}(\Mh) \rangle
\end{equation}
Another useful parameter in this context is the ratio of the two characteristic masses for hosting satellites and centrals, $M_{\rm 1}/M_{\rm min}$, which we plot as well below.

In Fig.~\ref{fig:fithod_g125}, we examine how these six parameters vary with environment, as measured by $\delta_{\rm 1.25}$, for the $n=0.01 \hmpcc$ galaxy sample.  The black dots in the left-hand panel are the measured halo occupation functions for centrals and satellites in the sample (the same as the black dotted and dashed curves in Fig.~\ref{fig:HOD125}). We fit the five-parameter model to the halo occupation function treating the centrals and satellites separately (solid curves). Following the choices made by \citet{Contreras2017} and \citet{Zehavi2018}, the fits assume equal weight to all measurements and use only mass bins with $\langle N(\Mh)\rangle>0.01$. We also neglect the most massive bin of $\sim 10^{15} \Msun$ which contains only a few haloes.  
We assign error bars by satisfying $\chi^2/{\rm d.o.f.}=1$.  We note that it is hard to fit well the turn-over of the centrals occupation around $\langle N(\Mh)\rangle=1$. The specific parameter values are shown as the black horizontal lines in the small panels on the right-hand side. 

To study the dependence of HOD on the environment, we fit the five-parameter HOD for each individual 10 per cent subsample of the ranked environment property. The HODs for just the two extreme subsets, corresponding to the 10 per cent highest and lowest ranked $\delta_{\rm 1.25}$, are shown as well on the left side of Fig.~\ref{fig:fithod_g125} in red and blue, respectively. The dots are again the occupation function calculated directly for the galaxies in these subsets of haloes, and the coloured solid lines are the HOD best fits. It is interesting to note that the satellites occupation exhibits a slightly curved shape (and not a pure power law), particularly noticeable for the upper 10 per cent $\delta_{\rm 1.25}$. Fig.~\ref{fig:fithod_g125} also shows the best-fit values of the individual HOD parameters, for each ranked 10 per cent subset.   These are represented by the green points with error bars in the small panels on the right-hand side, reflecting the variation of the HOD parameters with ranked $\delta_{\rm 1.25}$. The green curve is a spline fit to these points.  We clarify that the two HODs shown on the left-hand side of the figure for the lowest and highest 10 per cent of $\delta_{1.25}$ correspond to the left-most and right-most parameter values, respectively, in each small panel.

We see that the parameters' dependence on environment is complex.  While the slope $\alpha$ stays roughly constant, all other parameters vary to some degree. Of particular note is the large variation in the $M_{\rm 1}/M_{\rm min}$ ratio, which decreases with increasing density. In more detail, for the centrals occupation parameters, ${\rm log}M_{\rm min}$ depends weakly on $\delta_{\rm 1.25}$ for low densities and then varies for higher densities. We can consider ${\rm log}M_{\rm min}$ to broadly be decreasing with density, which is consistent with the overall trend that haloes in denser regions start hosting centrals at lower halo mass. The scatter ${\sigma_{{\rm log}M}}$ in general slightly increases with density, which implies a ``softer'' transition from unoccupied to occupied haloes.  There is some interplay between these two parameters, such that to first order the change with density can be described by either one (see \S~\ref{sec:edhod} below). The dependence of the satellite mass parameters on environment is relatively more distinct. Both $M_1$ and $M_{\rm cut}$ show a clear decrease with increasing $\delta_{\rm 1.25}$, which agrees with the OV observed. We note that the decreasing trend in both these parameters is similar in amplitude and slope.  Again, the satellites power-law slope parameter $\alpha$ experiences a very weak dependence on $\delta_{\rm 1.25}$.  While both of $M_{\rm min}$ and $M_1$ decrease with $\delta_{\rm 1.25}$, the latter trend is more pronounced, resulting in the significant decrease of the $M_1/M_{\rm min}$ ratio with $\delta_{\rm 1.25}$. 

Our results above for the dependence of the HOD parameters on $\delta_{\rm 1.25}$ are consistent with those found in \citet{Zehavi2018} for $\delta_{\rm 5}$ (their fig.~7), once accounting for the different smoothing scale. We also extend this investigation to the tidal anisotropy property $\alpha_{\rm 0.3, 1.25}$ in Appendix~\ref{sec:appendix}, and find that the parameter dependences are similar, but with a smaller amplitude.

\subsection{Mock catalogues with GAB}
\label{subsec:mock}

In \S~\ref{subsec:gab-env} we determined that the halo environment, such as that measured by $\delta_{\rm 1.25}$, can capture the bulk of the GAB signature. Building on the OV and HOD fits for the environment dependence investigated in \S~\ref{subsec:hodparam}, we now proceed to produce mock catalogues which mimic realistic GAB.  We do this by populating haloes with models that include the environmental dependence and test their ability to reproduce the GAB signal. We first create mock catalogues according to a simple interpolation scheme which incorporates the OV of a given secondary property. We also construct mock catalogues based on the inferred HOD parameters. The aim and importance of this is to demonstrate that indeed these properties can capture and produce the full level of GAB.  For illustrative purposes, we also do the same with halo concentration, $c$, which fails in reproducing the bulk of the GAB signature.

We begin with mock catalogues that directly utilize the measurements of the OV in the different subsamples of ranked environment property and interpolate between them to obtain the HOD for any ranked value. The advantage of such mocks is that they are free from any assumption regarding the HOD shape. Assuming a parametrized form for the occupation function, while convenient and captures the essential features, carries its own limitations by virtue of the restricted shape of the halo occupation function.  As we saw in \S~\ref{subsec:hodparam}, certain aspects of the occupation functions deviate from this restricted shape, which might introduce uncertainties into any mock catalogue based on this form. Hence, we begin our investigation with a parameter-free interpolation method.

For the ``interpolation mocks'', we start with the OVs of the secondary property which we have measured in ranked bins of 5 per cent. We tried cases with either 10, 20, or 30 bins of the secondary property, finding that 20 bins work best. Increasing the number of bins will not improve the performances of the mock catalogues. Again, we use directly the occupation function measurements and the HOD parametrization is not assumed. The number of galaxies inside any given halo is determined by interpolating the occupation numbers between two adjacent ranked bins of secondary property according to the rank of the halo in consideration. The detailed steps of creating the interpolating mock catalogue are as follows: \begin{enumerate} 
    \item Split haloes in each mass bin into 20 smaller bins according to the rank of secondary property $x$, such as 0-5 per cent, 5-10 per cent, and so on. Measure $\langle N_{\rm cen}(\Mh,x) \rangle$ and $\langle N_{\rm sat}(\Mh,x) \rangle$ in these rank bins, and consider them as the centrals and satellite occupations for haloes with $x$ rank of 2.5, 7.5 per cent, etc. (i.e. the center of the rank bins);
    \item For a given halo, obtain the specific rank of $x$ in its halo mass bin. Then interpolate between the adjoining bins to calculate $\langle N_{\rm cen}(\Mh,x)\rangle$ and $\langle N_{\rm sat}(\Mh,x)\rangle$ for the specific rank of the halo;
    \item Add Bernoulli scatter to $\langle N_{\rm cen}(\Mh,x)\rangle$ and Poisson scatter to $\langle N_{\rm sat}(\Mh,x)\rangle$ to obtain the actual central occupation $N_{\rm cen}(\Mh,x)$ and the number of satellites $N_{\rm sat}(\Mh,x)$ in the halo;
    \item If $N_{\rm cen}(\Mh,x)$=1, put the central galaxy at the centre of the halo (ie. the location of the most bound particle) and set the satellites according to step below. If $N_{\rm cen}(\Mh,x)$=0, we also set $N_{\rm sat}(\Mh,x)$=0;
    \item For $N_{\rm sat}(\Mh,x)$>0, assign the radial positions according to an NFW profile \citep{NFW}. To produce an NFW profile for each halo, we adopt Eq.~ 20 in \citet{Klypin2016} to assign the concentration $c$ according to $\vmax/\vvir$.  The angular positions of the satellites with respect to the central galaxy are assigned randomly. We caution the reader that as the satellite galaxies in the SAM do not follow exactly an NFW profile, the small-scale clustering will vary somewhat \citep{Jimenez2019}, however that does not impact the assembly bias effects measured on large scales.
\end{enumerate}

\begin{figure*}
	\centering
	\begin{subfigure}[h]{0.48\textwidth}
        \includegraphics[width=\textwidth]{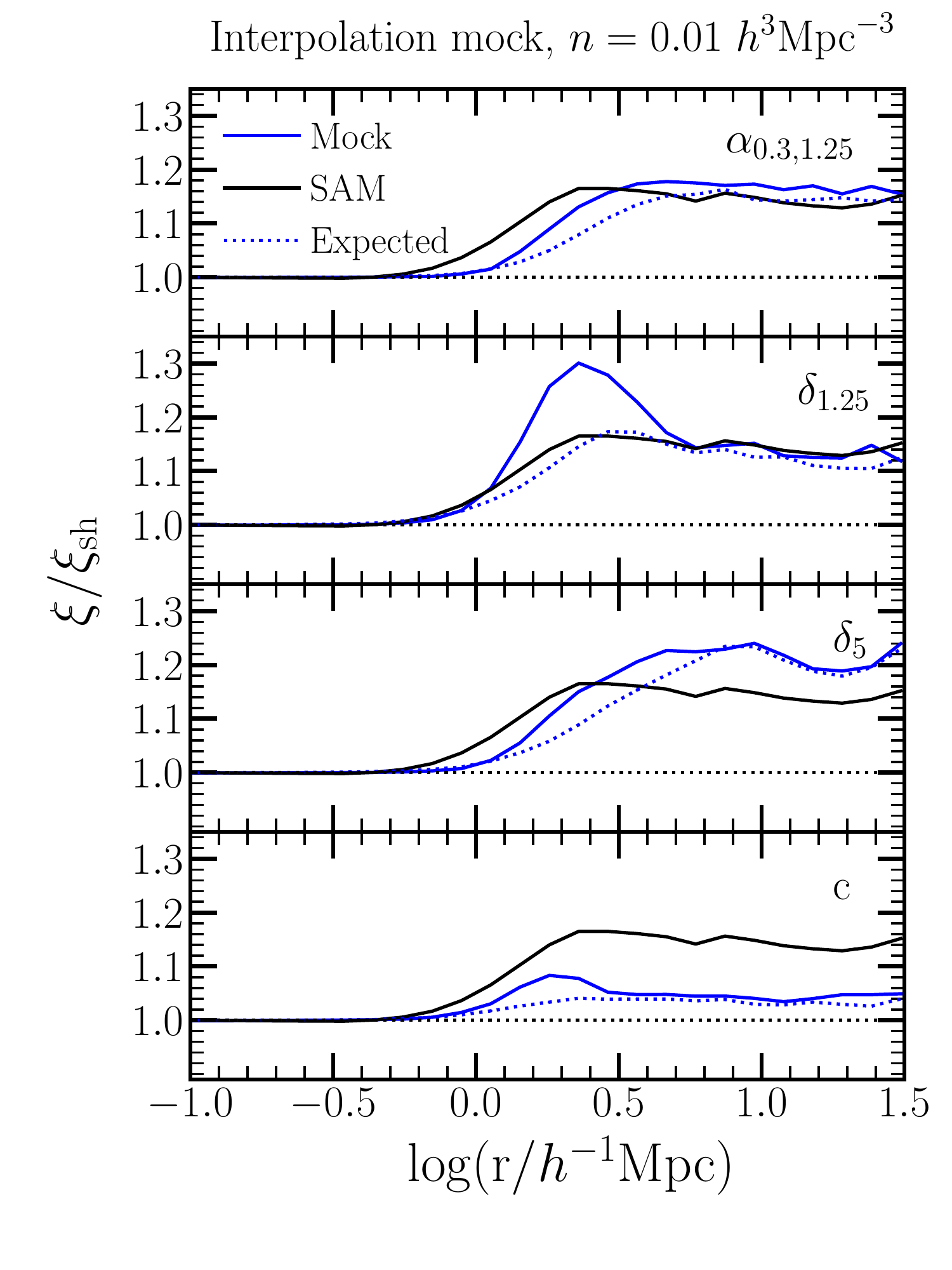}
	\end{subfigure}
	\hfill
	\begin{subfigure}[h]{0.48\textwidth}
		\includegraphics[width=\textwidth]{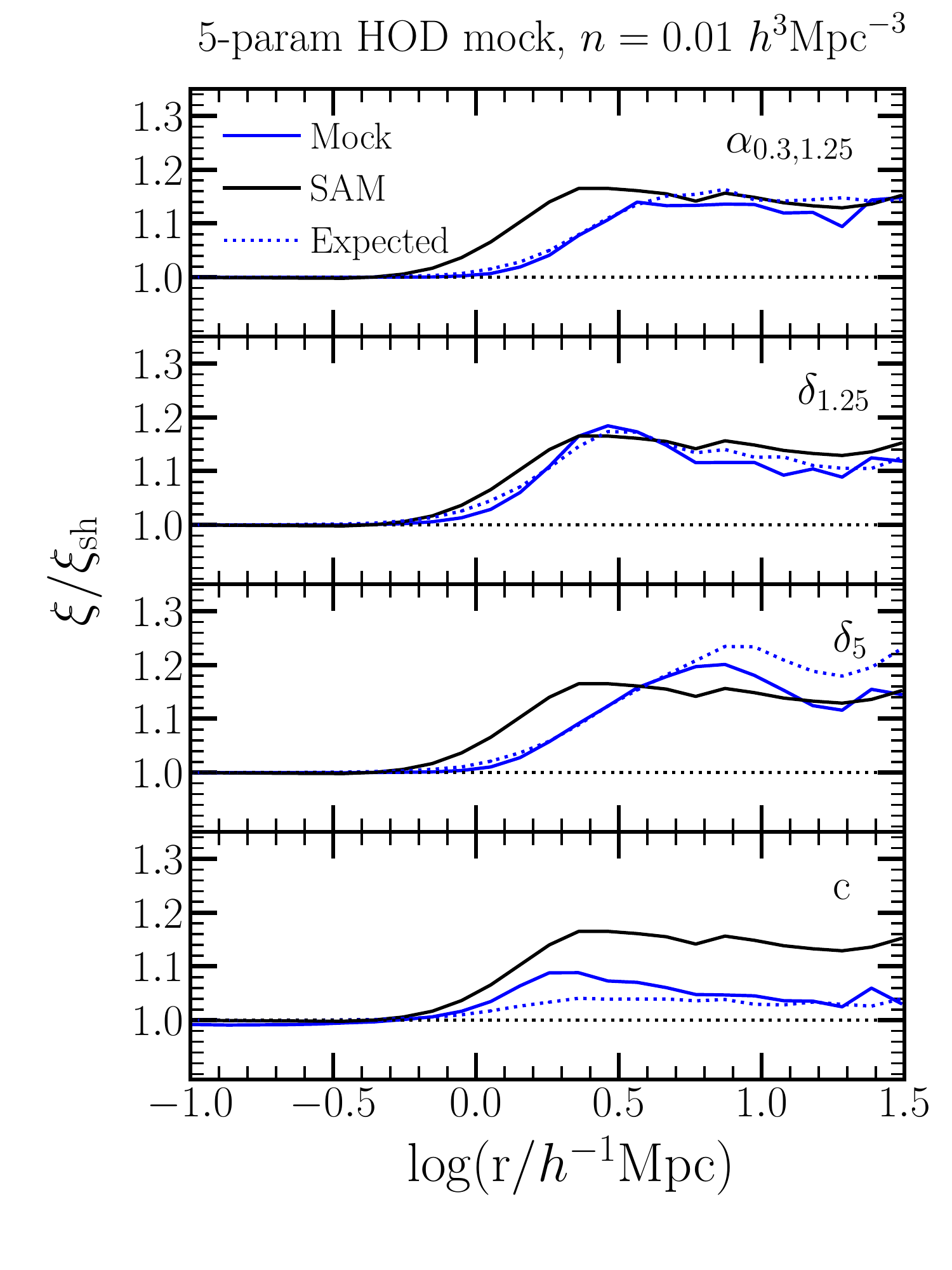}
	\end{subfigure}
	\hfill
\caption{GAB levels of the interpolation mock catalogues (left) and the 5-parameter HOD mock catalogues (right). These mock catalogues are based on $\alpha_{\rm 0.3,1.25}$, $\delta_{\rm 1.25}$, $\delta_{\rm 5}$, and concentration as described in the text. The black solid line, identical in all panels, represents the total level of GAB in the original galaxy sample, calculated as the ratio of the galaxy correlation function to that of the shuffled-by-mass galaxy sample. The blue solid line in each panel shows the level of GAB captured by the mock catalogues based on the different halo/environment property, obtained in the same manner. Also shown, for reference, as dotted blue lines are the results of the shuffling test for each of the individual properties used (these are the same as the respective coloured lines in Fig.~\ref{fig:shuffle_inner} and Fig.~\ref{fig:xienv}).
}
\label{fig:mock_hodfit}
\end{figure*}

We proceed to create such mock catalogues based on $\alpha_{\rm 0.3, 1.25}$, $\delta_{\rm 1.25}$, $\delta_{\rm 5}$, and the concentration parameter $c$, and measure the GAB level incorporated in each by measuring the ratio of the correlation function of the mock galaxy sample to that of a mass-only shuffled mock sample. The left-hand side Fig.~\ref{fig:mock_hodfit} shows our results for the $n=0.01 \hmpcc$ galaxy samples in the interpolation mock catalogues.  The solid blue lines are the GAB level in the mocks based on the different properties, while the solid black lines (identical in all panels) are the ``expected'' GAB level measured in the original SAM galaxy sample.  For comparison, we also mark (as dotted blue lines) the corresponding amounts of GAB associated with these specific secondary properties as estimated in \S~\ref{sec:gab}.

We find, for all secondary properties examined, that the GAB level on large scales in the interpolation mock catalogues (solid blue lines) is consistent with the amount of GAB associated with the same secondary property in SAM (dotted blue lines). This is to be expected, but is still an important proof of concept that one can take individual secondary properties, create a mock catalogue incorporating their OV, and recover their respective level of GAB.  For $\delta_{\rm 1.25}$ and $\alpha_{\rm 0.3,1.25}$, the GAB in the mock catalogue is at comparable levels as the full GAB in the SAM (black solid line). Hence, using these properties, one can in fact incorporate the correct full level of GAB into synthetic galaxy catalogues.

We note that, for the $\delta_{\rm 1.25}$ case, the clustering in the mock catalogue overestimates the shuffled result at the 1-halo to 2-halo transition regime ($\sim 2 \hmpc$). Such a``bump'' feature is also found (with differing amplitude) in other GAB studies (e.g. \citealt{Zehavi2018,Contreras2019,Contreras2020}). It is likely related to differences in the satellites distribution or the splashback galaxies in the outskirts of haloes impacting this transition regime. While we are unsure of the exact cause for it, we are mostly interested here in the clustering on large scales (above $\sim 5 \hmpc$), and we do not expect it to impact our conclusions regarding GAB.

We have included in this analysis also $\delta_{\rm 5}$, the dark matter density measured with a larger 5 $\hmpc$ Gaussian smoothing,  as it is often considered to probe the large-scale environment (e.g. \citealt{Zehavi2018}). In that case, the GAB in the mock sample is somewhat higher than the full SAM GAB, in accordance with the GAB produced in the shuffling test examined in \S~\ref{sec:gab} when holding halo mass and $\delta_{\rm 5}$ fixed. We will see later that this can still be ``tuned down'' and utilized to produce realistic mock catalogues. In contrast, for concentration $c$, which only accounts for a small fraction ($\sim$20 per cent) of the full SAM GAB, we see that the mock catalogue based on it also reproduces only $\sim$20 per cent of the full GAB. These results explain the difficulties of using concentration, or other internal halo properties for that matter, to incorporate assembly bias into mock catalogues (e.g. S.\ McLaughlin, in prep.; \citealt{Hadzhiyska2020a}). 

The second type of mock catalogues we consider is based on the OV quantified in terms of the  HOD parameters, inferred from the environmental percentiles in \S~\ref{subsec:hodparam}. We refer to it hereafter as the five-parameter HOD mock catalogues. The basic idea is to obtain the specific five parameters of the HOD for the environmental rank of the halo in consideration by interpolating between the parameters of the percentile rank bins. We tried different number of bins in this case as well, finding that 10 bins (i.e. 10 percentile each) work best. The steps of creating this mock catalogue for a secondary property $x$ are as follows:

\begin{enumerate} 
    \item Split haloes in each mass bin into 10 smaller bins according to the rank of secondary property $x$, such as 0-10 per cent, 10-20  per cent, and so on. Measure $\langle N_{\rm cen}(\Mh,x) \rangle$ and $\langle N_{\rm sat}(\Mh,x) \rangle$ in these rank bins;
    \item Fit the five-parameter HOD model to each of the $\langle N_{\rm cen}(\Mh,x) \rangle$ and $\langle N_{\rm sat}(\Mh,x) \rangle$ measurements to determine the values of the HOD parameters, and consider them as the parameters for haloes with $x$ rank bins centred on 5 per cent, 15 per cent, etc.;
    \item For a given halo, obtain the rank of $x$ in its halo mass bin. Calculate the five HOD parameters for this $x$ rank by interpolating the HOD parameter values between the values of the adjoining bins.  This provides the central and satellite mean occupation for this halo using Eq.~\ref{eq:cen} and Eq.~\ref{eq:sat};
    \item Follow the same final three steps as for the previous kind of mock catalogue to add scatter to $\langle N_{\rm cen}(\Mh,x)\rangle$ and $\langle N_{\rm sat}(\Mh,x)\rangle$ and assign galaxies to the individual haloes.
\end{enumerate}

We present the results for the 5-parameter HOD mock catalogues in the right-hand side of Fig.~\ref{fig:mock_hodfit}, for the same halo properties as the other set of mocks.  We compare the GAB measurements obtained for these mock samples (solid blue lines) with the GAB measured in the original SAM $n=0.01 \hmpcc$ galaxy sample (the same solid black line). For $\alpha_{\rm 0.3,1.25}$ we essentially recover the same level of GAB as in the original sample, and a nearly identical measure as the one obtained from the double shuffling test (dotted blue line). For $\delta_{\rm 1.25}$ and $\delta_{\rm 5}$, the recovery is similar to that of the original sample, but there appears to be a small shift in the amplitude of the recovered GAB relative to their expected level from the earlier shuffling test.  The reason for this offset is likely the restrictive shape assumed by this HOD parametrization that doesn't agree in detail when fitting the individual rank bins. This results in the GAB level lying slightly below the expected level for the  $\delta_{\rm 1.25}$ case, and for the predicted GAB in the $\delta_{\rm 5}$ case to now agree better with the original full level of GAB.  For the concentration, the situation remains similar to that of the interpolation mock catalogues, with the GAB recovered by it amounting to only $\sim$20 per cent of the full level. It is interesting to note that the ``bump'' feature on intermediate scales is largely suppressed in the 5-parameter HOD mocks.

With the results above, we find that the mock catalogues which incorporate the OVs directly or the HOD parameter dependence on the secondary property can generally reproduce the same level of GAB as that associated with the same property in the SAM. In particular, if this secondary property is $\delta_{\rm 1.25}$ or $\alpha_{\rm 0.3,1.25}$, the resulting synthetic samples have the same amount of GAB on large scales as that of the original galaxy sample. In the below section, we propose a simple modification of the traditional five-parameter HOD to incorporate the environmental dependence and produce the proper level of GAB.

\section{Modified HOD based on environment}
\label{sec:edhod}

The mock catalogues described in \S~\ref{subsec:mock}, based on the variations with one environmental property, are generally successful in incorporating the correct level of GAB.  While they serve as an important proof of concept, their application is somewhat cumbersome and specific to the SAM in hand. Motivated by their success, we now proceed with a simple extension of the HOD model that incorporates the environment dependence.  The result is a modified HOD model which is both practical and tunable to any GAB level.

\subsection{Modified 7-parameter HOD model}
\label{subsec:7paramHOD}

The standard five-parameter HOD \citep{Zheng2005} is by construction assembly-bias free, and thus does not capture the more complex galaxy-halo connection resulting in GAB.  In Fig.~\ref{fig:fithod_g125}, we have demonstrated the dependence of the five HOD parameters on environment as captured in the SAM. Here we present an extension of the traditional HOD model by taking a secondary property into consideration. Instead of modifying the HOD parameters as a function of the actual value of the secondary property, we focus on the rank of the secondary property. This allows us to probe the dependence across the whole halo mass range, factoring out the correlation of halo mass with environment, and allows for a simpler modelling. By shifting to rank values, we are also less sensitive to the specific values and exact definition of the environment. The secondary property can be any environmental property that can represent the correct level of GAB, and we use $\delta_{\rm 1.25}$ as a detailed example. In Appendix~\ref{sec:appendix2} we show the results for $\delta_{\rm 5}$ and $\alpha_{0.3,1.25}$ as well.

We first assign each halo a ranked value of $\delta_{1.25}$ between 0 and 1, in narrow bins of halo mass (these are the same values quoted as percentiles in Fig.~\ref{fig:fithod_g125}).  Relative to the median value of 0.5, ranks above it correspond to overdense regions and ranks below it represent underdense environments.  Generally, all five standard HOD parameters (Eq.~\ref{eq:cen} and \ref{eq:sat}) may depend on $\delta_{\rm 1.25}$ to different degrees, as can be seen in Fig.~\ref{fig:fithod_g125}. We start with the values of these parameters for the mass-only HOD, and aim for the simplest modification that can capture the main changes of these parameters and recover the right level of GAB.  To that effect we introduce two additional parameters, $B_{\rm cen}$ and $B_{\rm sat}$, that quantify the level of OV present in the central and satellite occupations.

Given the fundamental role that $M_{\rm min}$ and $M_{\rm 1}$ have as the two characteristic halo masses for hosting central galaxies and satellites, our proposed modification is as follows:
\begin{equation}
\label{eq:Mminmod}
      {\rm log}M_{\rm min}(\delta_{\rm 1.25}^{\rm rank})={\rm log}M_{\rm min}^0 + B_{\rm cen} \times [\delta_{\rm 1.25}^{\rm rank} - 0.5] 
\end{equation}
and
\begin{equation}
\label{eq:M1mod}  
      {\rm log}M_{\rm 1}(\delta_{\rm 1.25}^{\rm rank})={\rm log}M_{\rm 1}^0 + B_{\rm sat} \times [\delta_{\rm 1.25}^{\rm rank} - 0.5] \,.  
\end{equation}
$M_{\rm min}^0$ and $M_{\rm 1}^0$ are the values of the standard mass-only HOD parameters,  which by definition are also the values for the median $\delta_{\rm 1.25}^{\rm rank}$. The sign of the assembly bias parameters, $B_{\rm cen}$ and $B_{\rm sat}$, signifies the sense of the trend with environment, with a negative value corresponding to the mass scales decreasing with increased $\delta_{\rm 1.25}$. The absolute values of these parameters indicate the maximal range (in dex) over which these logarithmic mass scales vary with environment. A larger value corresponds to a larger OV and generally a higher level of GAB.

For the centrals HOD parameters, we choose to modify $M_{\rm min}$, in response to the trend of galaxies preferentially occupying haloes in dense environments, causing the shift of the centrals occupation function toward lower halo mass. As shown in Fig.~\ref{fig:fithod_g125}, both $M_{\rm min}$ and $\sigma_{{\rm log}M}$ vary somewhat with $\delta_{1.25}$, so in principle either of these parameters could have been chosen (or both). However, we tested modifying $\sigma_{{\rm log}M}$ while holding $M_{\rm min}$ fixed, and found that it can not reproduce the centrals GAB in the SAM on its own. Additionally, the observational constraints on $\sigma_{{\rm log}M}$ are less robust. Since we aim for a simple extension of the HOD model with one additional central parameter, it is reasonable to proceed with $M_{\rm min}$.
With regard to the three satellite parameters (Eq.~\ref{eq:sat}),  $M_{\rm cut}$ and $M^*$ change in a similar manner with $\delta_{\rm 1.25}$,  while the slope $\alpha$ remains nearly constant (as also shown in Fig.~\ref{fig:fithod_g125}).  Equation~\ref{eq:M1mod} can be rewritten as $M_1(\delta_{\rm 1.25}^{\rm rank})=M_1^0 10^{B_{\rm sat}[\delta_{\rm 1.25}^{\rm rank}-0.5]}$, which effectively translates to similar expressions for $M_{\rm cut}$ and $M^*$ with the same $B_{\rm sat}$ coefficient: 
\begin{equation}
\label{eq:Mcutmod}  
      {\rm log}M_{\rm cut}(\delta_{\rm 1.25}^{\rm rank})={\rm log}M_{\rm cut}^0 + B_{\rm sat} \times [\delta_{\rm 1.25}^{\rm rank} - 0.5] \,,  
\end{equation}
and
\begin{equation}
\label{eq:M*mod}  
      {\rm log}M^*(\delta_{\rm 1.25}^{\rm rank})={\rm log}M^{*0} + B_{\rm sat} \times [\delta_{\rm 1.25}^{\rm rank} -0.5] \,.  
\end{equation}
One can surely generalize this model to include changes to the other parameters if needed. However, we show below that these two assembly bias parameters are adequate to describe the clustering properties of the SAM.

Other extensions of the HOD to incorporate assembly bias have been proposed in the literature in recent years.  \citet{Paranjape2015} correlate galaxy colours to the halo concentration, resulting in a tunable halo model of galactic conformity.  \citet{Hearin2016} present a more general framework to incorporate a secondary dependence (often the halo concentration) into the HOD. \citet{Yuan2018,Yuan2020} employ a similar modification of the HOD, also using the halo concentration in a step-wise manner. \citet{Hadzhiyska2020b} consider a two-dimensional HOD that augments the model with a secondary parameter in addition to mass. In a different approach, more similar to ours, \citet{McEwen2018} extend the traditional HOD form to include parameters based on the mass density environment. Based on their analysis of the age-matching mock catalogues of \citet{Hearin2013}, they vary only the central occupation function, such that ${\rm log}M_{\rm min}$ and ${\sigma_{{\rm log}M}}$ depend linearly on the value of density field. Following this work, \citet{Wibking2019} and \citet{Salcedo2020} utilize a model in which $M_{\rm min}$ is modified as a function of the large-scale environment, and the $M_1/M_{\rm min}$ ratio is either held fixed or allowed to vary, determining the change in the satellite occupation. 

While envisioned independently, our own work follows the same spirit of these latter papers.  Similarly to these works, we use the ranked value of $\delta$ as a measure that can be used across all halo masses. We choose to focus on a smaller smoothing scale ($1.25 \hmpc$ Gaussian) because of its success in reproducing the full level of GAB (with either $\delta$ or $\alpha_{\rm 0.3}$, as shown in \S~\ref{subsec:gab-env}), however, we also explore this modelling with the larger $5 \hmpc$ Gaussian smoothing in Appendix~\ref{sec:appendix2}.  In contrast with these works \citep{McEwen2018,Wibking2019,Salcedo2020},  we model independently the variation of the satellites occupation with environment, given their distinct occupancy variation and the significant dependence of $M_1/M_{\rm min}$ on environment as shown in Fig.~\ref{fig:fithod_g125}. In what follows, we produce mock catalogues using this 7-parameter HOD model and confirm that it is able to capture the full level of GAB. 

\subsection{Mock catalogues with modified HOD model}
\label{subsec:7paramHOD-mock}

\begin{figure*}
	 \includegraphics[width=1.0\textwidth]{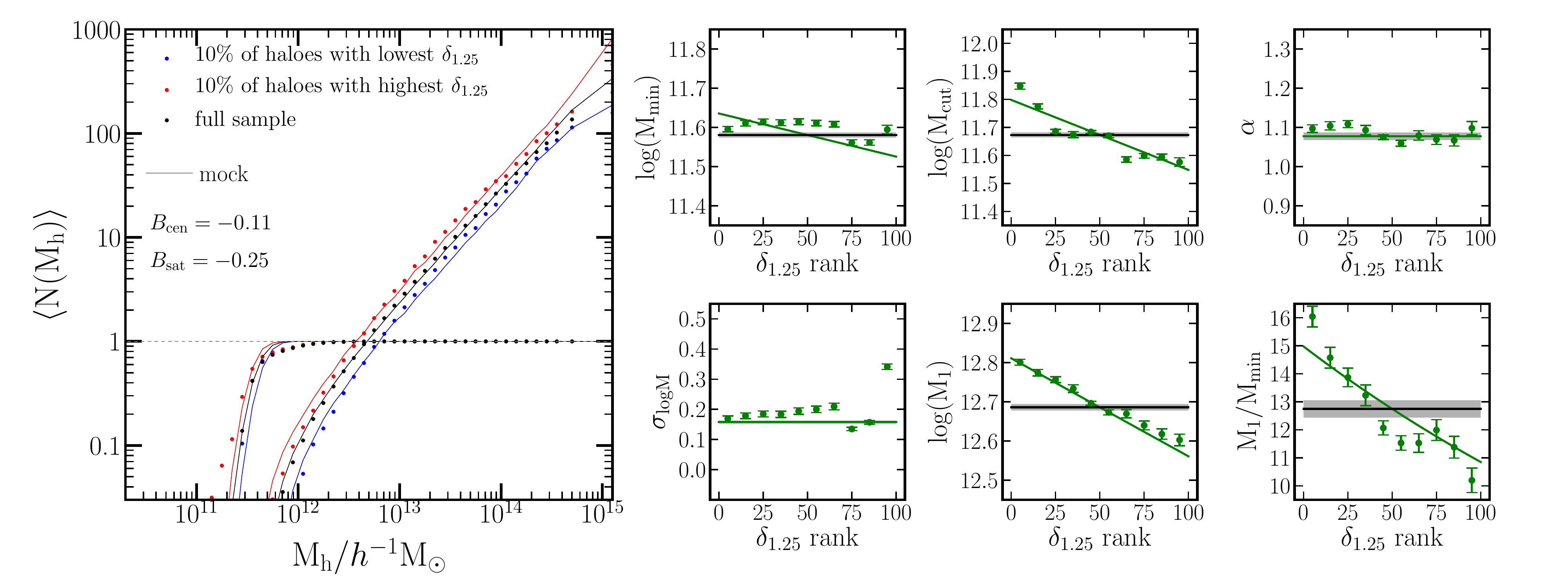}
\caption{
Halo occupation functions and corresponding parameters as a function of $\delta_{\rm 1.25}$ for the mock catalogue using the modified 7-parameter HOD representing the $n=0.01 \hmpcc$ number density. Dots in all panels are the measurements from the SAM catalog, the same as shown in Fig.~\ref{fig:fithod_g125}. In the left-hand side, the black curves are the central and satellites occupation functions of the full galaxy sample in the mock catalogue.  The red and blue curves show the occupation functions for the mock galaxies in the 10 per cent of haloes with highest and lowest values of $\delta_{\rm 1.25}$, respectively. The green solid lines in the 6 small panels indicate the corresponding modified HOD model, while the black solid represents the standard HOD parameters from the full SAM sample.
}
\label{fig:mHODparam_g125}
\end{figure*}

We now proceed to create mock catalogues based on this modified model, and fit the assembly bias parameters $B_{\rm cen}$ and $B_{\rm sat}$ by comparing the resulting level of GAB to that measured in the SAM catalogue. We first measure the clustering of the central galaxies only in order to determine $B_{\rm cen}$, and then add satellites to obtain the value of $B_{\rm sat}$ from the clustering of the full sample. Instead of a formal chi-square fitting, we start with an initial guess of the parameters, create the corresponding mock catalogue and compare the GAB levels, adjust the parameter accordingly and repeat. The detailed steps of fitting the additional parameters are as follows:
\begin{enumerate}
    \item Obtain the five standard HOD parameters $M_{\rm min}$, $\sigma_{{\rm log}M}$, $M_{\rm cut}$,$M^*$, and $\alpha$ from fitting to the halo occupation function of the full sample. Assume an initial guess for $B_{\rm cen}$; 
    \item Rank the $\delta_{\rm 1.25}$ values in bins of halo mass, and associate a rank value to each halo.
    \item For each halo, use this rank to obtain the modified central parameter ${\rm log}M_{\rm min}(\delta_{\rm 1.25}^{\rm rank})$ for the initial $B_{\rm cen}$ according to Eq.~\ref{eq:Mminmod}. Create a mock catalogue with only central galaxies using the modified HOD parameter ${\rm log}M_{\rm min}(\delta_{\rm 1.25}^{\rm rank})$ and the unmodified parameter $\sigma_{{\rm log}M}$;
    \item Measure the clustering of the central galaxies and that of the shuffled sample to obtain the centrals GAB signal in the mock catalogue and compare to that of the SAM;
    \item If the centrals GAB level in the mock catalogue is higher/lower than that of the SAM, then lower/increase the absolute value of $B_{\rm cen}$ and repeat steps (iii) and (iv) until the mock centrals GAB matches the SAM central GAB ($f_{\rm AB}\sim1$);
    \item Proceed with an initial guess for $B_{\rm sat}$. Calculate the modified satellite parameters $M_{\rm cut}(\delta_{\rm 1.25}^{\rm rank})$ and $M^*(\delta_{\rm 1.25}^{\rm rank})$ for all haloes using Eq.~\ref{eq:Mcutmod} and \ref{eq:M*mod};
    \item Create a mock catalogue with central and satellite galaxies using the $B_{\rm cen}$ determined in step (v) and the assumed $B_{\rm sat}$. The locations and scatter of the mock galaxies are determined in the same manner as described in Section~\ref{subsec:mock}.
    \item  Measure the GAB in the mock catalogue and compare it to that of the SAM catalogue.
    \item If the mock GAB is higher/lower than that of the SAM, lower/increase the absolute value of $B_{\rm sat}$ and repeat step (vii) and (viii), until the mock GAB matches the SAM GAB.
\end{enumerate}

Doing this process for the $n=0.01 \hmpcc$ galaxy sample, using the $\delta_{\rm 1.25}$ ranks, results in values of $B_{\rm cen}=-0.11$ and $B_{\rm sat}=-0.25$ for the assembly bias parameters. As expected, both these parameters have negative values, corresponding to a decreased halo mass for larger densities. The absolute value of $B_{\rm sat}$ is significantly larger than that of $B_{\rm cen}$, implying a stronger OV for the satellites, as can also be inferred from Fig.~\ref{fig:HOD125}.

Fig.~\ref{fig:mHODparam_g125} (left-hand side) shows the occupancy variations in the mock sample corresponding to this model and the change of the standard HOD parameters.  On the left-hand side, we plot the central and satellites occupation functions for the full galaxy sample in the mock (black curves), and for the galaxies in the 10 per cent of the haloes with the highest (red) and lowest (blue) values of $\delta_{1.25}$. The dots are the corresponding direct measurements in the SAM.  We can see that the bulk of the occupancy variations is captured by this model.  The remaining small differences between the mock and SAM measurements are in fact at the same level of the differences exhibited in the analogous panel of Fig.~\ref{fig:fithod_g125}, where each subset was fitted by the five-parameter model separately.  Such differences are to be expected due to the constrained shape of the HOD model.  Thus the OV level of $\delta_{\rm 1.25}$ in the mock sample is approximately the same as that in the SAM. However, since this modified model only depends on $\delta_{\rm 1.25}$, it is not able to reproduce the OV of other properties that are not correlated tightly with $\delta_{\rm 1.25}$, like the concentration and age.

The right-hand side of Fig.~\ref{fig:mHODparam_g125} shows the values of the individual standard HOD parameters and their variation with $\delta_{1.25}$. The green dots are the values measured directly in the SAM (identical to the ones shown in Fig.~\ref{fig:fithod_g125}), while the green lines show our modified seven-parameter model used in creating the mock sample. As discussed above, the values of $\sigma_{{\rm log}M}$ and $\alpha$ remain fixed at the value obtained for the full sample, while all other parameters depend linearly on the rank of $\delta_{\rm 1.25}$. The slope is set by either $B_{\rm cen}$ (for ${\rm log}M_{\rm min}$) or $B_{\rm sat}$ (for ${\rm log}M_{\rm cut}$ and ${\rm log}M_{1}$) as shown.   We see that the modified model follows the trend of the SAM reasonably well for the satellite parameters,  while the centrals variation is perhaps a bit more refined, but still captured in essence by our model.  It is reassuring that the significant variation of $M_{1}/M_{\rm min}$ with environment (which is not modelled independently but rather inferred) is reproduced well by our model.  We examined these diagnostics for the other number density galaxy samples as well, and find comparable agreement. In an analogous fashion, we also compute the modified seven-parameter HOD models based on the variations with $\delta_{\rm 5}$ and $\alpha_{\rm 0.3,1.25}$, and find similar results which are shown in Appendix~\ref{sec:appendix2}.

\begin{figure*}
	\centering
	\begin{subfigure}[h]{0.32\textwidth}
	\includegraphics[width=\textwidth]{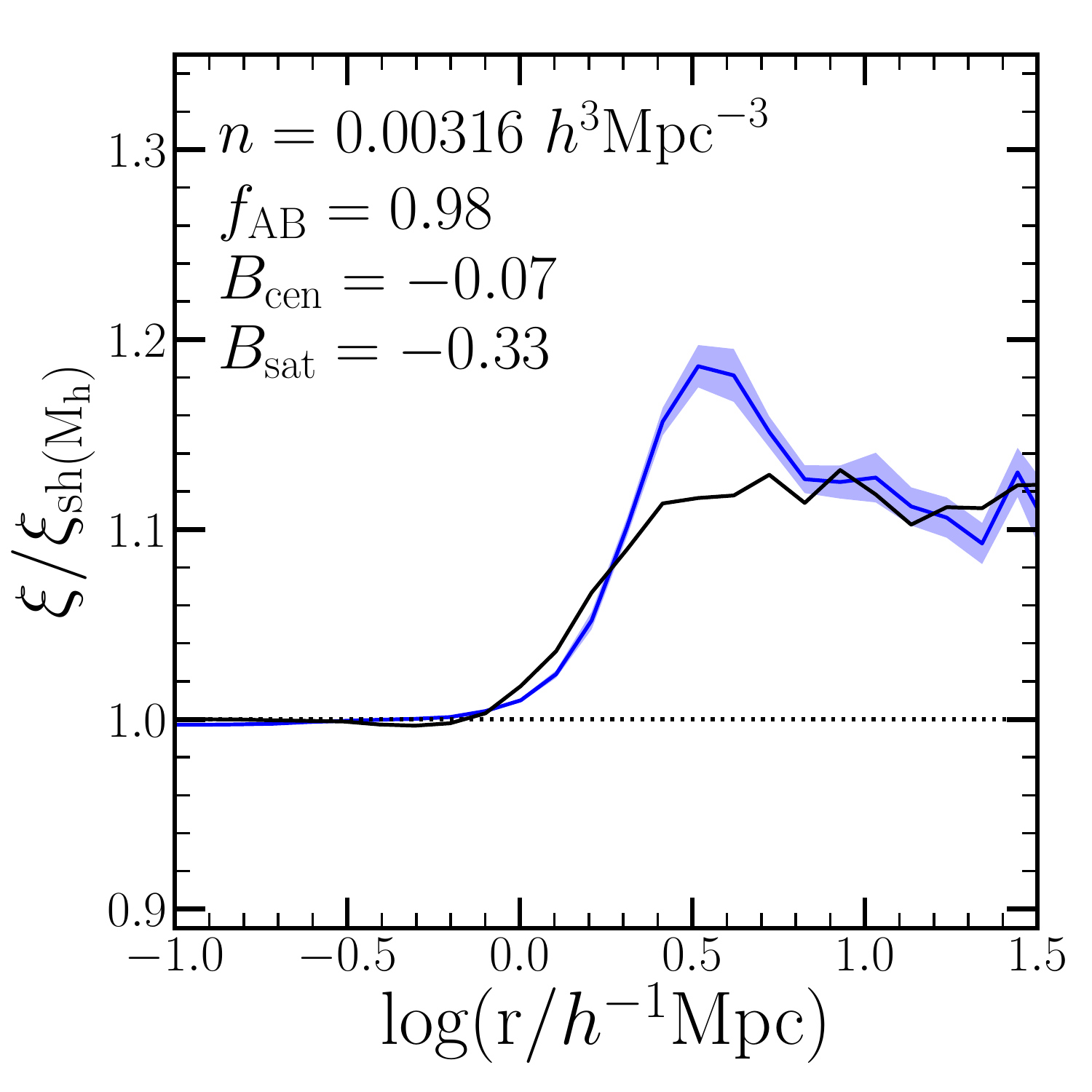}
	\end{subfigure}
	\hfill
	\begin{subfigure}[h]{0.32\textwidth}
        \includegraphics[width=\textwidth]{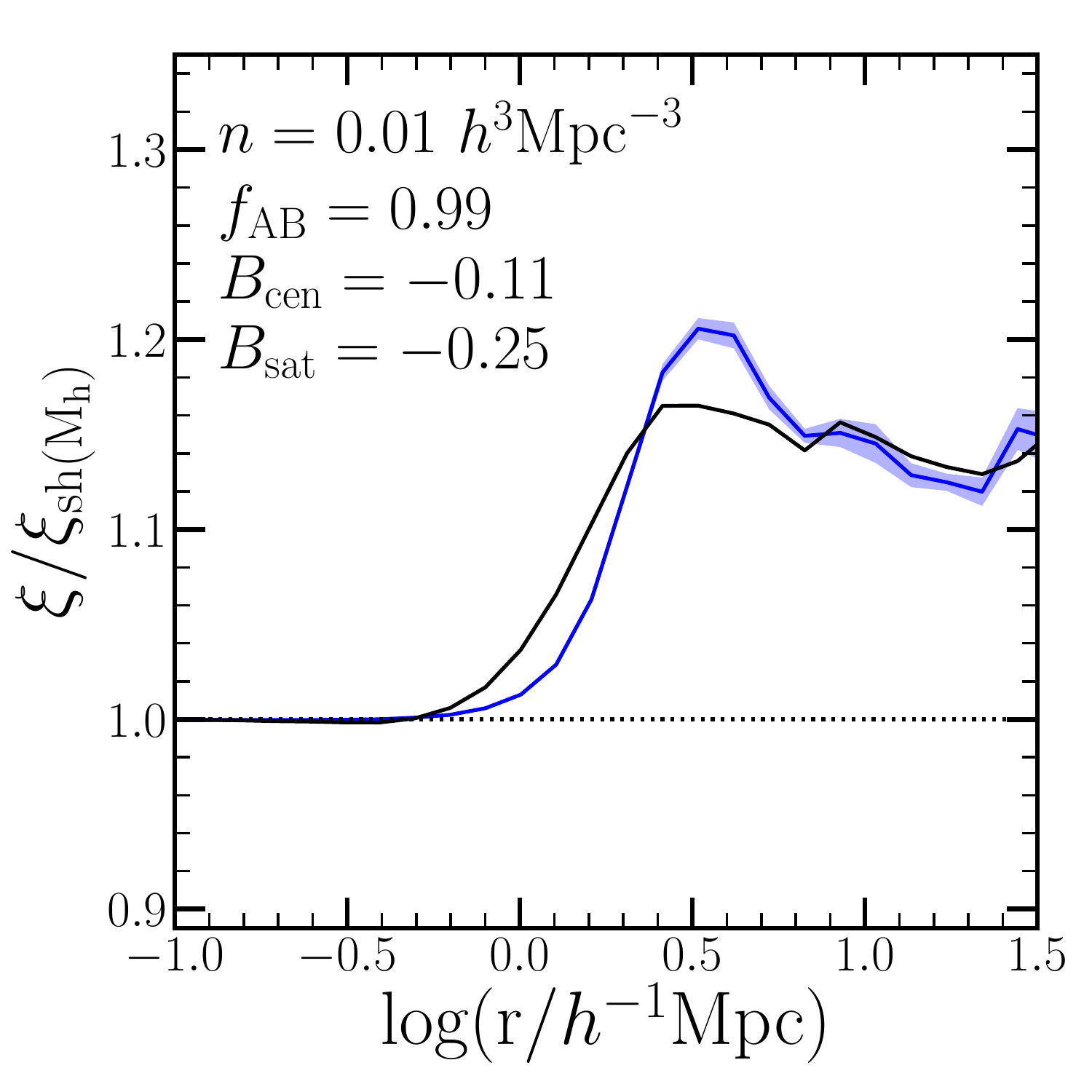}
	\end{subfigure}
	\hfill
	\begin{subfigure}[h]{0.32\textwidth}
        \includegraphics[width=\textwidth]{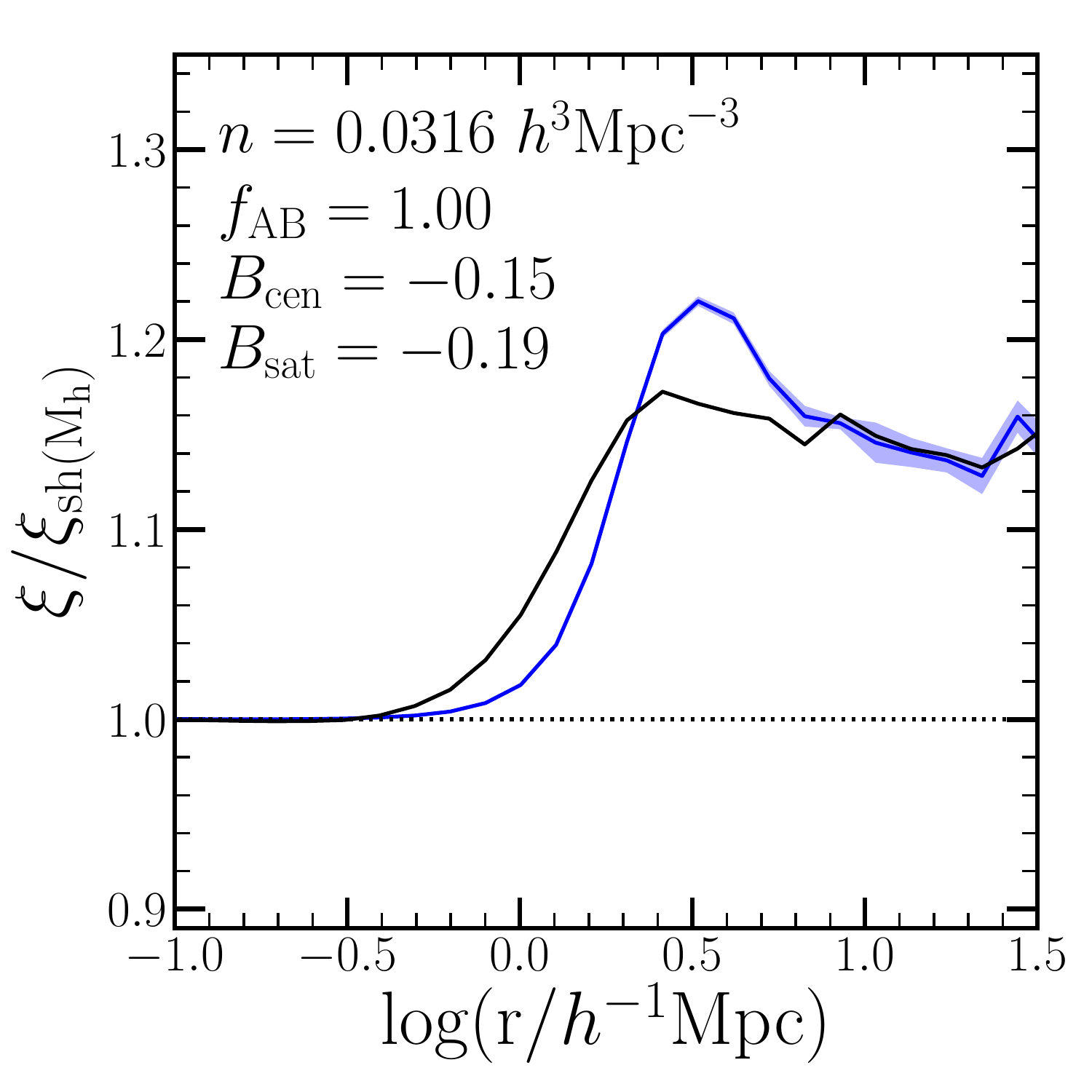}
	\end{subfigure}
	\hfill
\caption{GAB levels of modified HOD mock based on $\delta_{\rm 1.25}$. From left to right: GAB level of the mock galaxies (blue curve) and the SAM catalogue (black curve) for galaxy number densities $0.00316 \hmpcc$,  $0.01 \hmpcc$, and $0.0316 \hmpcc$. The shaded blue region represents the uncertainty on the mock GAB measurement from 10 different shufflings. The value of $f_{\rm AB}$ denotes the fraction of SAM GAB captured by the mock galaxies on large scales. Also specified are the values of the two additional assembly bias parameters used for each mock catalogue.
}
\label{fig:mHODgab_g125}
\end{figure*}

Finally, Fig.~\ref{fig:mHODgab_g125} presents the GAB levels of the mock samples for the modified HOD models which match the GAB in the SAM catalogue, shown here for the three galaxy number densities considered.  Specifically, the middle panel shows the results for the $n=0.01 \hmpcc$ sample analysed in Fig.~\ref{fig:mHODparam_g125}. The values of the assembly bias parameters in each case are labelled in the individual panels.  The ``bump'' feature appears here as well, likely due to differences in the satellites distribution that affect the 1-halo to 2-halo transition regime. These, however, do not impact the level of GAB measured on large scales.
The fraction of the SAM GAB reproduced, $f_{\rm AB}$, is marked in each figure. We see that for all number density samples, our simple modified HOD model is able to reproduce the full level of GAB. Additional fine-tuning of the parameters may reach a value of $f_{\rm AB}$ even closer to unity, however, the randomness involved with the shuffling mechanism (for both the mock samples and the SAM catalogue), as well as the scatter around the mean occupation, limit the accuracy of the fitting. This makes our measurements effectively indistinguishable from a full recovery of the GAB signal.

Overall, we see that our proposed seven-parameter HOD model encapsulates the environment dependence, such that a mock catalogue based on this model is able to reproduce the correct level of galaxy assembly bias.   To recap, our modified model has seven free parameters ($M_{\rm min}$, $\sigma_{{\rm log}M}$, $M_{\rm cut}$, $M^*$, $\alpha$, $B_{\rm cen}$, and $B_{\rm sat}$) that describe the occupation function of central galaxies and satellites as a function of both halo mass and environment.  In this work, we determine the new assembly bias parameters, $B_{\rm cen}$ and $B_{\rm sat}$, by first obtaining the five traditional parameters from the SAM and then setting them to match the GAB level. Alternatively, one can also fit the seven parameters simultaneously (in particular if the correct value of the standard five parameters is not known). This model can be used to produce mock catalogues that contain a specific level of GAB, matched to different galaxy formation models, and also to examine the GAB level in observational galaxy samples.   

\section{Summary and conclusion}
\label{sec:summary}
We investigate the importance of internal halo properties and environmental measures to galaxy assembly bias, using fixed number density samples defined by stellar mass, derived from the \citet{Guo2011} SAM galaxy formation model implemented on the Millennium simulation. To measure the total amount of GAB in the SAM, we compute the ratio of the correlation function of the original SAM sample to that of a shuffled sample, where the galaxy content of haloes is randomly reassigned to haloes of the same mass. To assess the contribution of individual secondary properties, we perform a double shuffling,  at fixed mass and fixed secondary property,  which effectively removes the impact of all other properties. We then examine the ratio of the clustering of this galaxy sample to that of the shuffled-by-mass sample, and compare it to the full level of GAB in the SAM.

We find that the internal halo properties account for only a small fraction of the full GAB. For example, the commonly used halo formation time and halo concentration contribute only 26 per cent and 21 per cent of the signal, respectively, for the $n=0.01 \hmpcc$ number density sample. The highest fraction is obtained for the number of subhaloes $n_{\rm sub}$ inside a halo, amounting to $\sim$30 per cent of the full GAB.  In contrast, environmental properties prove to be more important for GAB. This is perhaps not unexpected given that the correlation function and the environment measure different aspects of clustering,  but it is still insightful to explore in detail. The matter density measured with a $1.25 \hmpc$ Gaussian smoothing ($\delta_{\rm 1.25}$) is able to reproduce the full level of GAB, while larger smoothing scales result in an even increased signal. Measures of the tidal anisotropy also play an important role, but care has to be taken as to the exact definition and smoothing scale. The tidal anisotropy parameter $\alpha_{\rm 1, 1.25/2.5/5/10}$ corresponds to only a few per cent of the total GAB measurement, while a slightly different definition of $\alpha_{\rm 0.3, 1.25}$ can reach the full level. We conclude that the environmental properties $\delta_{\rm 1.25}$ and $\alpha_{\rm 0.3, 1.25}$ are the most informative for recovering the full GAB in the SAM, and thus can be used to create mock catalogues that include the correct level of assembly bias. Our finding that the environment measured on a relatively-small $1.25 \hmpc$ Gaussian smoothing scale is the most informative may be related to the turn around radius of halos (e.g. \citealt{Fong20}) and is compatible with related results for HAB (e.g. \citealt{Paranjape2018a,Han2019}).

As GAB is the combined result of halo assembly bias and the occupancy variations, we proceed to explore the OV with $\delta_{\rm 1.25}$ and $\alpha_{\rm 0.3, 1.25}$, which is helpful for creating mock samples and modifying the traditional HOD model. The OV with $\delta_{\rm 1.25}$ is similar to that of $\delta_{\rm 5}$ \citep{Zehavi2018}, such that haloes in denser regions tend to host central galaxies (above a stellar mass threshold) at lower halo mass, and have more satellites at fixed halo mass. The tidal anisotropy $\alpha_{\rm 0.3, 1.25}$ is correlated to $\delta_{\rm 1.25}$ and has similar OV (while $\alpha_{\rm 1, 1.25}$ shows an opposite trend).
In order to study in detail the dependence of the halo occupation on these environmental measures, we utilize the standard five-parameter HOD model and fit it to subsamples of the haloes grouped by the rank of the secondary property. We show that all HOD parameters vary with the environment to some degree, perhaps with the exception of the power-law slope of the satellites occupation.  We proceed to build mock catalogues by both simply interpolating the OV measurements for the different ranks of environment and by using the fitted HOD parameters to these.  For both methods, we find that mock samples based on either $\delta_{\rm 1.25}$ or $\alpha_{\rm 0.3, 1.25}$ reproduce $\sim$100 per cent GAB level, while mock samples based on internal halo properties such as the concentration recover only a small fraction, consistent with the fraction of the full GAB associated with it.

Finally, we propose a modification of the standard five-parameter HOD form to incorporate the dependence on environment, by introducing two additional parameters which describe the level of GAB in the centrals and satellites occupations. Our seven-parameter model assumes a linear dependence of the logarithmic values of $M_{\rm min}$ and $M_{\rm 1}$, the characteristic halo masses for hosting centrals and satellites, on the rank of the environmental property. We focus on $\delta_{\rm 1.25}$ in this work, but any halo or environment property that can represent the correct GAB level can be used.  We fit these parameters by creating mock catalogues accordingly and matching the level of GAB for central galaxies on their own and for the full sample.

Our resulting modified HOD provides a practical way to incorporate assembly bias into the HOD framework,  and is tunable to the GAB level of different galaxy formation models. The practical applications are two-fold. First, it can be utilized for producing mock catalogues which incorporate realistic levels of GAB. These are becoming increasingly important for the predictions, testing, and analysis of upcoming large galaxy surveys. Second, our methodology can be applied directly to observational data with the aim of inferring the level of galaxy assembly bias in the real Universe, which we leave for future work. The two assembly bias parameters $B_{\rm cen}$ and $B_{\rm sat}$ may depend on the specific galaxy samples or galaxy formation physics, as well as on cosmology. Our general flexible model allows for these variations. If assembly bias is significant, it will be important to include these parameters to prevent systematics when constraining cosmology with future surveys.

Our modification of the HOD is in the same spirit of the modelling used by \citet{Wibking2019} and \citet{Salcedo2020} to incorporate the environment dependence, but our model allows for an independent variation of the satellites occupation, and we also explore alternative parameters such as the tidal anisotropy. Such a modelling can also be conceivably utilized in the framework of the decorated HOD \citep{Hearin2016}. Based on our results, however, we caution against the commonly used step-wise concentration ``decoration'', and instead advocate a linear change with the density. 
Subhalo abundance matching (SHAM) is another approach to connect galaxies and dark matter (sub)haloes using a monotonic relation between a galaxy property and a specified halo property, like the maximum circular velocity or infall mass (e.g. \citealt{Conroy2006,Reddick2013,Guo2016}), which by its nature includes some level of assembly bias \citep{Zentner2014,Chaves2016,Lehmann2017}.
Most recently, \citet{Contreras2020} propose a flexible SHAM-based model for GAB, linking the galaxy property to the large-scale bias of the halo. Similar to our modified HOD model, they incorporate the GAB with two free parameters. However, SHAM requires the subhaloes information, which might not be available for very large cosmological simulations with limited resolution.  In contrast, our modified HOD model can be fine-tuned to any galaxy formation model and easily applied to larger simulations to obtain the appropriate distribution of galaxies, matching both the correct clustering and the right level of assembly bias. 

While the simple modified HOD we propose has clear advantages in creating mock catalogues and exploring GAB in observational data, we note that it can not capture the OV for all halo or environmental properties simultaneously, since only one secondary property is used in the model. So for example, while our model recovers the correct level of GAB and the OV with density or tidal anisotropy, it might not reproduce the OV with concentration. To address that and recover the full range of dependences involved, we are employing machine learning techniques to infer the intricate relations and accurately connect the galaxies to dark matter haloes (S.\ Kumar et al., in prep.).
While assembly bias remains a challenge for contemporary models of galaxy clustering and the galaxy-halo connection, our work here already provides considerable insight into the nature of this complex phenomenon.  It provides a practical way to produce galaxy mock catalogues that incorporate this effect, crucial for upcoming large surveys, and it facilitates the measurement of assembly bias in the real Universe.

\section*{Acknowledgements}
We thank Aseem Paranjape, Ravi Sheth, and Zheng Zheng for useful discussions and technical suggestions. We also thank Nelson Padilla for comments. We thank the anonymous referee for insightful comments that helped improve the presentation of the paper. XX and IZ acknowledge support by NSF grant AST-1612085. SC acknowledges the support of the ``Juan de la Cierva Formaci\'on'' fellowship (FJCI-2017-33816).

\section*{Data Availability}
The data underlying this article are available in GitHub at \url{https://github.com/xiaojux2020/SAMGAB}


\appendix
\section{Additional results regarding the tidal anisotropy}
\label{sec:appendix}

\begin{figure}
	 \includegraphics[width=0.48\textwidth]{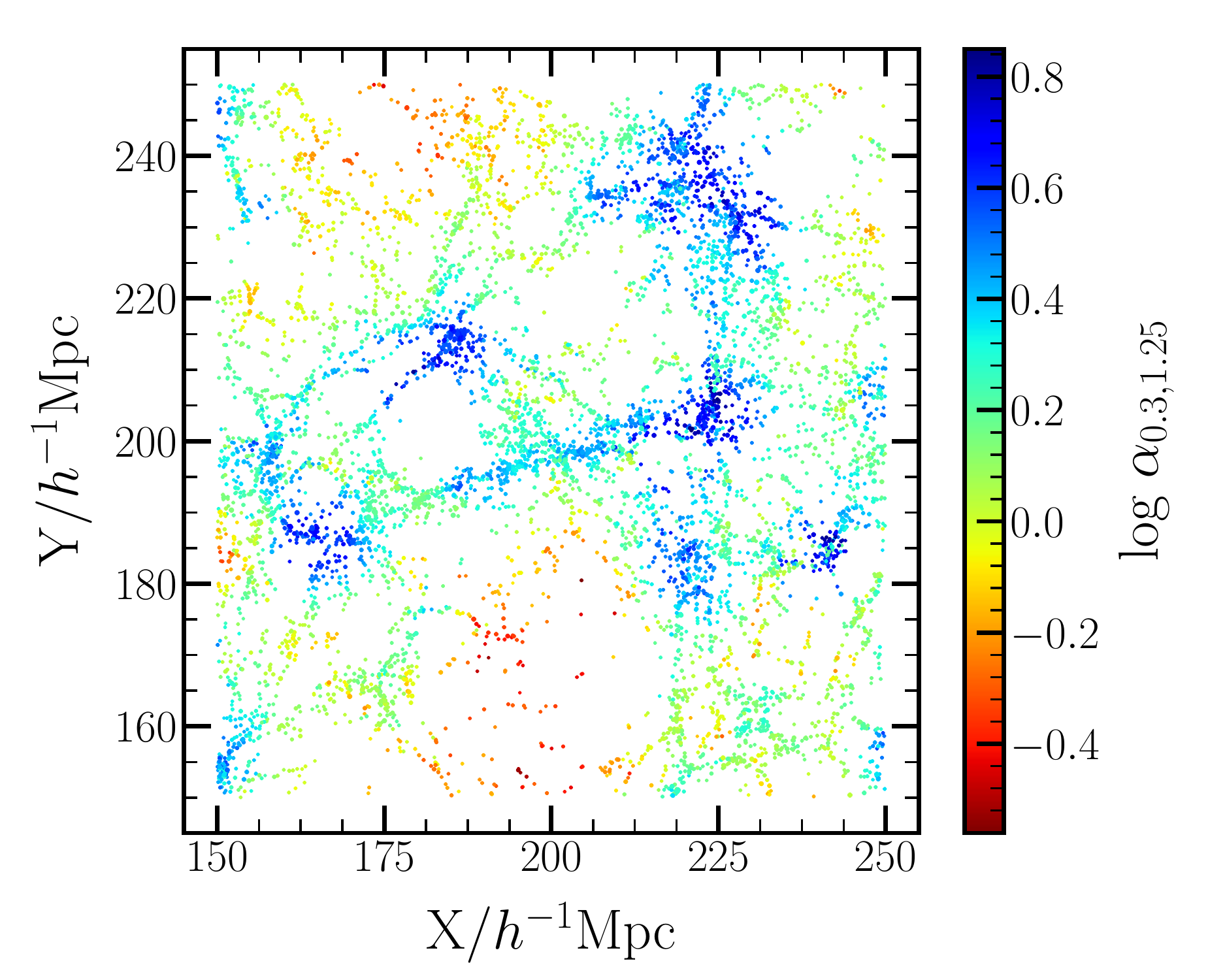}
\caption{
Spatial distribution of haloes in a $100 \hmpc \times 100 \hmpc\, \times 10 \hmpc$ slice of the Millennium simulation. The individual haloes are colour-coded according to the value of $\alpha_{\rm 0.3,1.25}$.
}
\label{fig:slice-alpha03}
\end{figure}

\begin{figure*}
	 \includegraphics[width=0.78\textwidth]{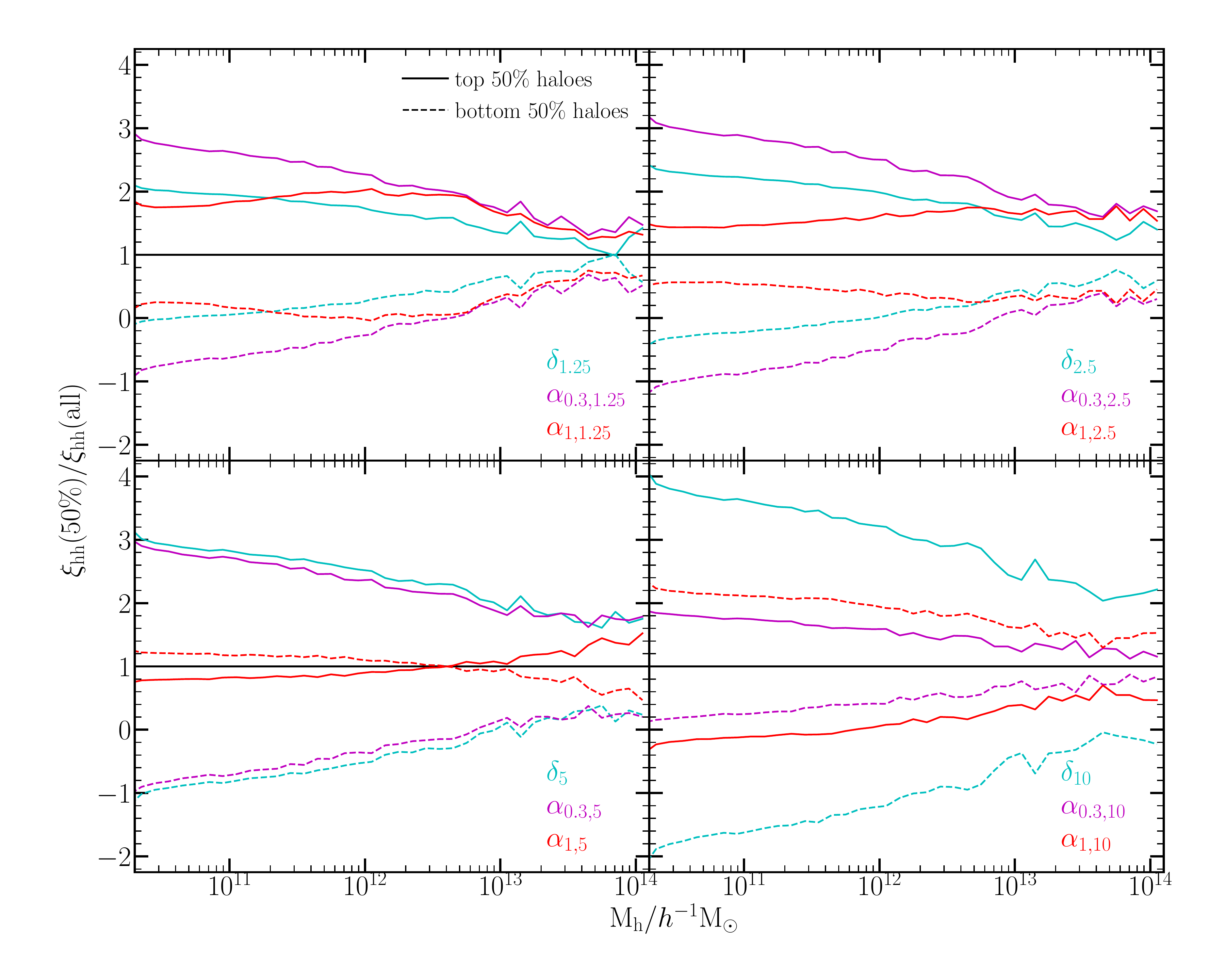}
\caption{
Cross-correlation functions of haloes with different environmental selections relative to that of the full sample as a function of halo mass. The environmental properties shown here are $\delta$ (cyan), $\alpha_{\rm 0.3}$ (magenta), and $\alpha_{\rm 1}$ (red). The solid coloured curves show the clustering ratios measured for the 50 per cent of haloes with the highest values for these properties, and the dashed curves show the ratios measured for the 50 per cent of haloes with the lowest values. Each panel corresponds to a different smoothing scale, 1.25, 2.5, 5, and 10 $\hmpc$, going from upper left to lower right, as labelled. The black solid line in each panel indicates the clustering of the full halo sample (ratio of unity). 
}
\label{fig:HABalpha03}
\end{figure*}

\begin{figure*}
    \includegraphics[width=1.0\textwidth]{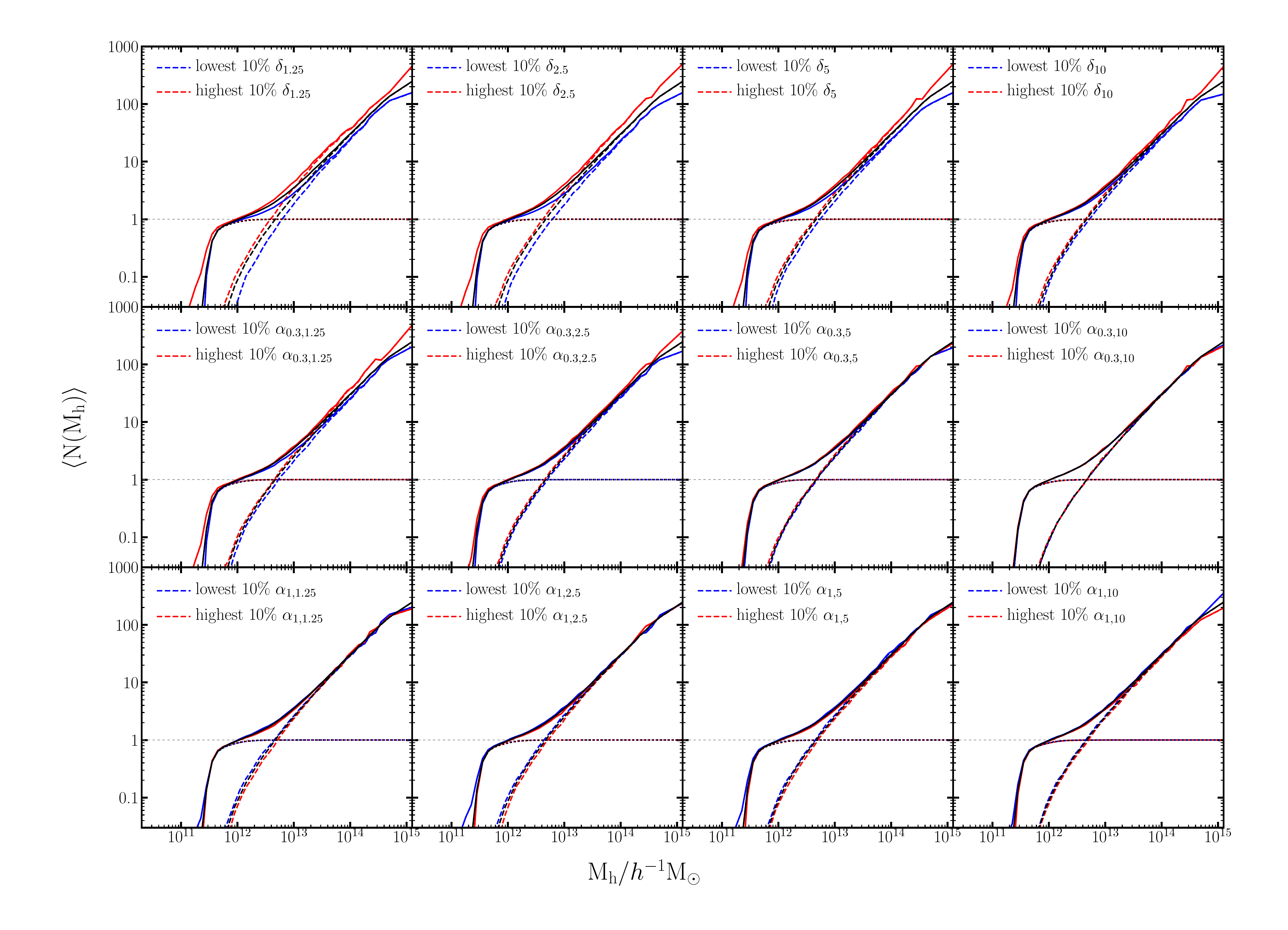}
    \caption{Halo occupation functions for the $n=0.01 \hmpcc$ galaxy sample, showing the occupancy variations with $\delta$ (top row), $\alpha_{\rm 0.3}$ (middle row) and $\alpha_{\rm 0.1}$ (bottom row). The different columns correspond to the four different smoothing scales, 1.25, 2.5, 5, and 10 $\hmpc$, increasing from left to right. In each panel,  the black line is the occupation of the full sample, while the red and blue lines correspond to the subset of galaxies in the 10 per cent of haloes with the highest and lowest values, respectively, of the secondary property. This is the same as shown in Fig.~\ref{fig:HOD125} for the 1.25 $\hmpc$ smoothing, but now including all smoothing scales and $\alpha_{\rm 1}$.
}
\label{fig:HODall}
\end{figure*}

In \S~\ref{subsec:gab-env}, we present the definition of the tidal anisotropy, and explore the galaxy assembly bias attributed to tidal anisotropy parameters such as $\alpha_{0.3}$ and $\alpha_{1}$, as well as to the different density measures. In \S~\ref{subsec:hod}, we also show the occupancy variation associated with $\alpha_{0.3,1.25}$ and $\delta_{\rm 1.25}$.  The GAB is in essence the convolution of the OV effects with those of halo assembly bias. For completeness, we present here the HAB associated with the tidal anisotropy and density measures as well as a set of the OV measurements.

We start by examining in Fig.~\ref{fig:slice-alpha03} the spatial distribution of haloes with mass lower than $10^{12.5}\hmsun$ in a slice of the simulation box colour coded by $\alpha_{\rm 0.3,1.25}$. Blue (red) dots correspond to haloes with higher (lower) values $\alpha_{\rm 0.3,1.25}$. It appears that haloes with the highest $\alpha_{\rm 0.3,1.25}$ values (blue) reside in nodes, haloes with intermediate values of $\alpha_{\rm 0.3,1.25}$ (green and yellow) tend to reside in the filamentary structures, and haloes with the lowest values (orange and red) are in the voids. This spatial distribution is consistent with that in \citet{Paranjape2018a}. There are also some similarities to the distribution of haloes coded by density (for example, as shown in \citealt{Zehavi2018}, fig.~1).  Hence, we expect regions of high $\alpha_{\rm 0.3,1.25}$ to generally also be regions of high density,  while capturing additional information about the tidal torque.

Fig.~\ref{fig:HABalpha03} presents the halo clustering differences for the different environment measures. We show the ratio of cross-correlation functions for the half of the haloes with the highest value of the environmental properties (solid coloured curves) and the half of the haloes with the lowest values (dashed coloured curves),  relative to the auto-correlation function of all haloes. These ratios are computed and shown as a function of halo mass, obtained from the ratios of the corresponding cross-correlation function and the full auto-correlation function averaged over large scales ($10-25 \hmpc$). The environmental properties considered are $\delta$ (cyan), $\alpha_{\rm 0.3}$ (magenta), and $\alpha_{\rm 1}$ (red). Each panel corresponds to a different smoothing scale as marked.  We drop the last few bins with halo mass above $\sim 10^{14}\Msun$, where the halo count falls below 100. While not explicitly a measure of halo bias in this case, the relative level of HAB can be directly inferred from the difference between the two lines, at fixed halo mass. 

Strikingly, the level of HAB for $\alpha_{\rm 0.3}$, for all smoothing scales, is very significant, with larger $\alpha_{\rm 0.3}$ corresponding to a larger bias. The level of HAB remains roughly similar for the smaller smoothing scales, but decreases for the $10 \hmpc$ case, while the HAB for $\delta$ increases monotonically with the smoothing scale. We see that the HAB with $\alpha_{\rm 0.3}$ is in fact stronger than that of $\delta$ for the two smallest smoothing scales. For the 5 $\hmpc$ smoothing scale, the HAB of $\alpha_{\rm 0.3}$ is slightly lower than that of $\delta$, and is much more so for the 10 $\hmpc$ case.
The behaviour of HAB for $\alpha_{\rm 1}$ is different. It is generally smaller than for $\alpha_{\rm 0.3}$, in particular for small halo masses.  In contrast to the other measures in Fig.~\ref{fig:HABalpha03}, the HAB for $\alpha_{\rm 1}$ continuously decreases with smoothing scale, inverting at about the 5 $\hmpc$ scale and continuing so. For the largest smoothing scale,  the HAB for $\alpha_{\rm 0.3}$ has in fact a larger amplitude than that of $\alpha_{\rm 1}$, but is in the {\it opposite} sense, such that haloes with smaller $\alpha_{\rm 1}$ values correspond to a larger bias.

It is reasonable to infer that for a smoothing scale smaller than 1.25 $\hmpc$ Gaussian, the HAB associated with $\alpha_{\rm 1}$ is increased and potentially larger than that of $\delta$, consistent with the result of \citet{Ramakrishnan2019} that the HAB dependence on $\alpha_{\rm 1}$, with variable smoothing of $4R_{\rm 200}$ top-hat, is more important than that on $\delta$ on that scale. It is clear that in certain regimes, the tidal anisotropy is associated with a very significant HAB, leading to considering it as a primary indicator of HAB. We have confirmed that other internal halo properties like concentration (not shown here) have a much reduced HAB in comparison, in accordance 
with the findings of \citet{Ramakrishnan2019}. We caution, however, that care must be taken as the signal is dependent on the exact definition of the tidal anisotropy parameter and the smoothing scale utilized, as demonstrated here.

In \S~\ref{subsec:hod}, we analysed the OV with $\delta_{\rm 1.25}$ and $\alpha_{\rm 0.3,1.25}$, for the $n=0.01 \hmpcc$ galaxy sample.  In Fig.~\ref{fig:HODall} we extend that to including all four smoothing scales as well as the tidal anisotropy variant $\alpha_{\rm 1, 1.25}$. The top left-hand panel for $\delta_{\rm 1.25}$ and the middle left-hand panel for $\alpha_{\rm 0.3, 1.25}$ are the cases already shown in Fig.~\ref{fig:HOD125}. We see that for higher values of these parameters, central galaxies start occupying lower mass haloes and the satellites occupation also shifts slightly toward lower masses.
For $\alpha_{\rm 1, 1.25}$, in contrast, we find the opposite trend (bottom left-hand panel), with the lower values of the tidal anisotropy corresponding to a shift toward lower mass scales.  The reason for this is the anti-correlation between $\alpha_{\rm 1, 1.25}$ and $\delta_{\rm 1.25}$, while $\alpha_{\rm 0.3,1.25}$ and $\delta_{\rm 1.25}$ is positively correlated. It is this somewhat unexpected behaviour of the OV for $\alpha_{\rm 1}$ that is likely leading to this parameter not contributing significantly to the GAB as explored in \S~\ref{subsec:gab-env}, and caused us to explore alternative definitions.
For all of the three variables in Fig.~\ref{fig:HODall}, the OV trends are weaker for larger smoothing scale. This is easy to understand, since the larger smoothing scales ``smear'' more the fields such that the haloes are considered to be in more similar environments.

\section{Additional results regarding the modified HOD}
\label{sec:appendix2}

\begin{figure*}
	 \includegraphics[width=1.0\textwidth]{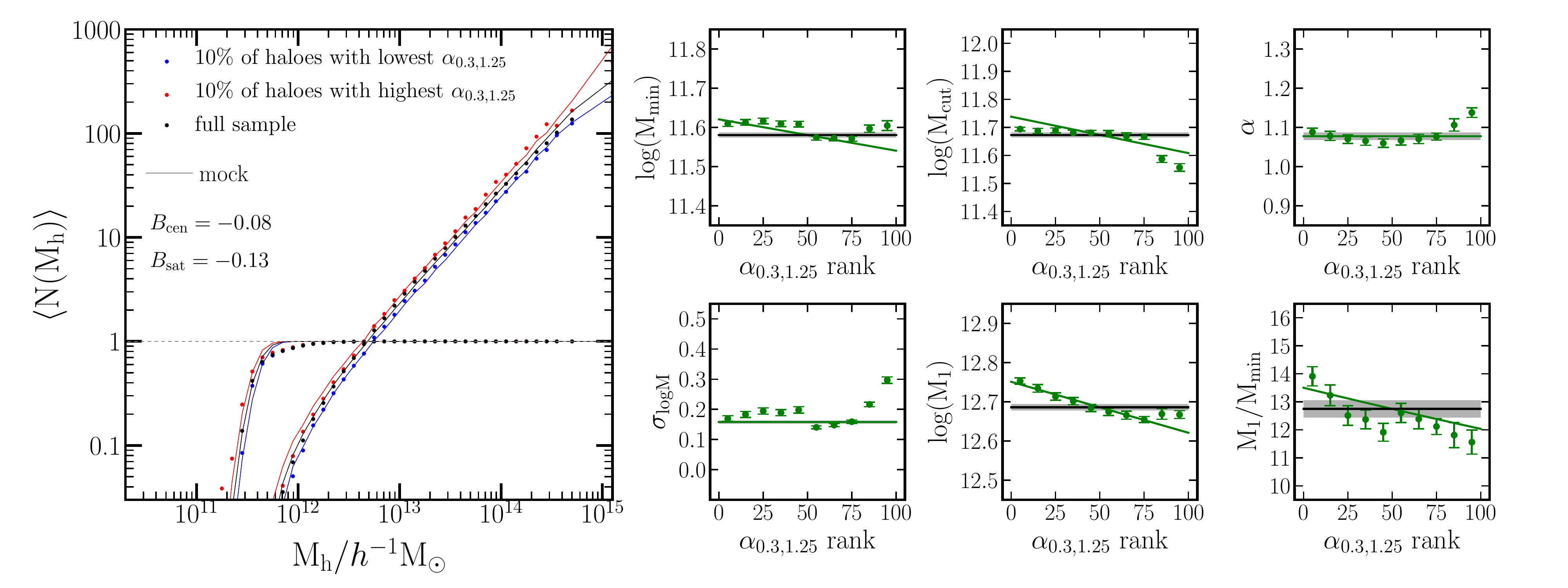}
\caption{
OV and HOD parameters of modified HOD mock sample based on $\alpha_{\rm 0.3,1.25}$. The same as
Fig.~\ref{fig:mHODparam_g125}, but for $\alpha_{\rm 0.3, 1.25}$.
}
\label{fig:mHODmock_alpha03g125}
\end{figure*}

\begin{figure*}
	 \includegraphics[width=1.0\textwidth]{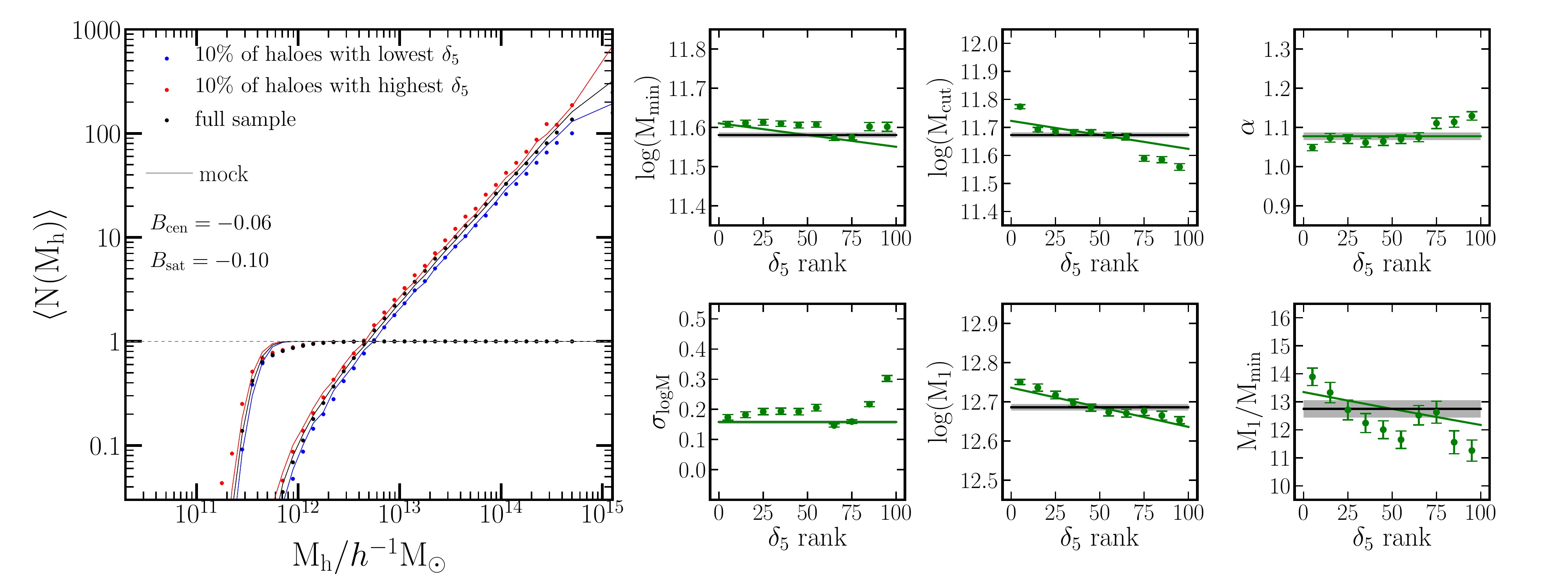}
\caption{
OV and HOD parameters of modified HOD mock sample based on $\delta_{\rm 5}$. The same as Fig.~\ref{fig:mHODparam_g125} and \ref{fig:mHODmock_alpha03g125} but for $\delta_{\rm 5}$.
}
\label{fig:mHODmock_g5}
\end{figure*}

\begin{figure}
	\centering
	\begin{subfigure}[h]{0.235\textwidth}
	  \includegraphics[width=\textwidth]{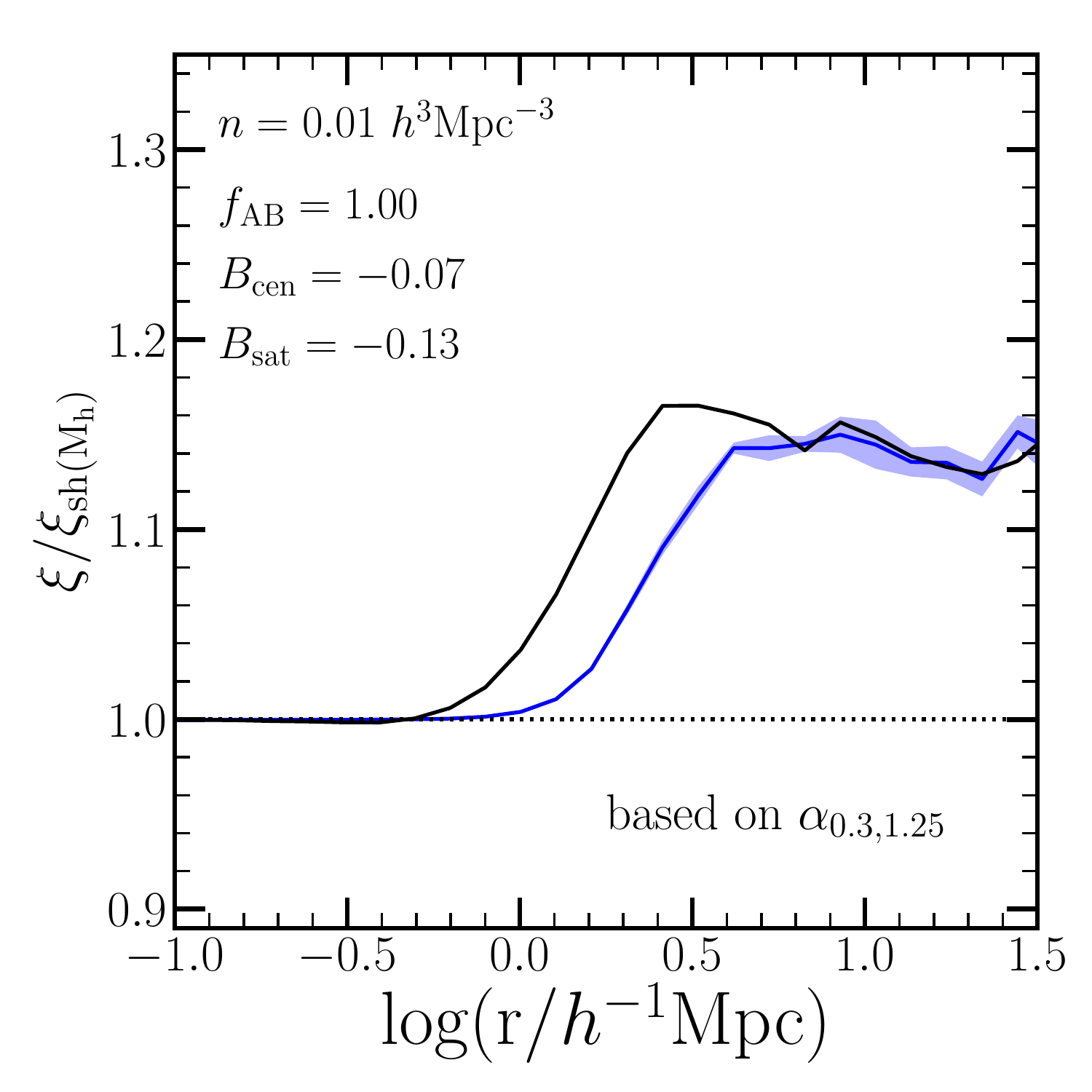}
	\end{subfigure}
	\hfill
	\begin{subfigure}[h]{0.235\textwidth}
          \includegraphics[width=\textwidth]{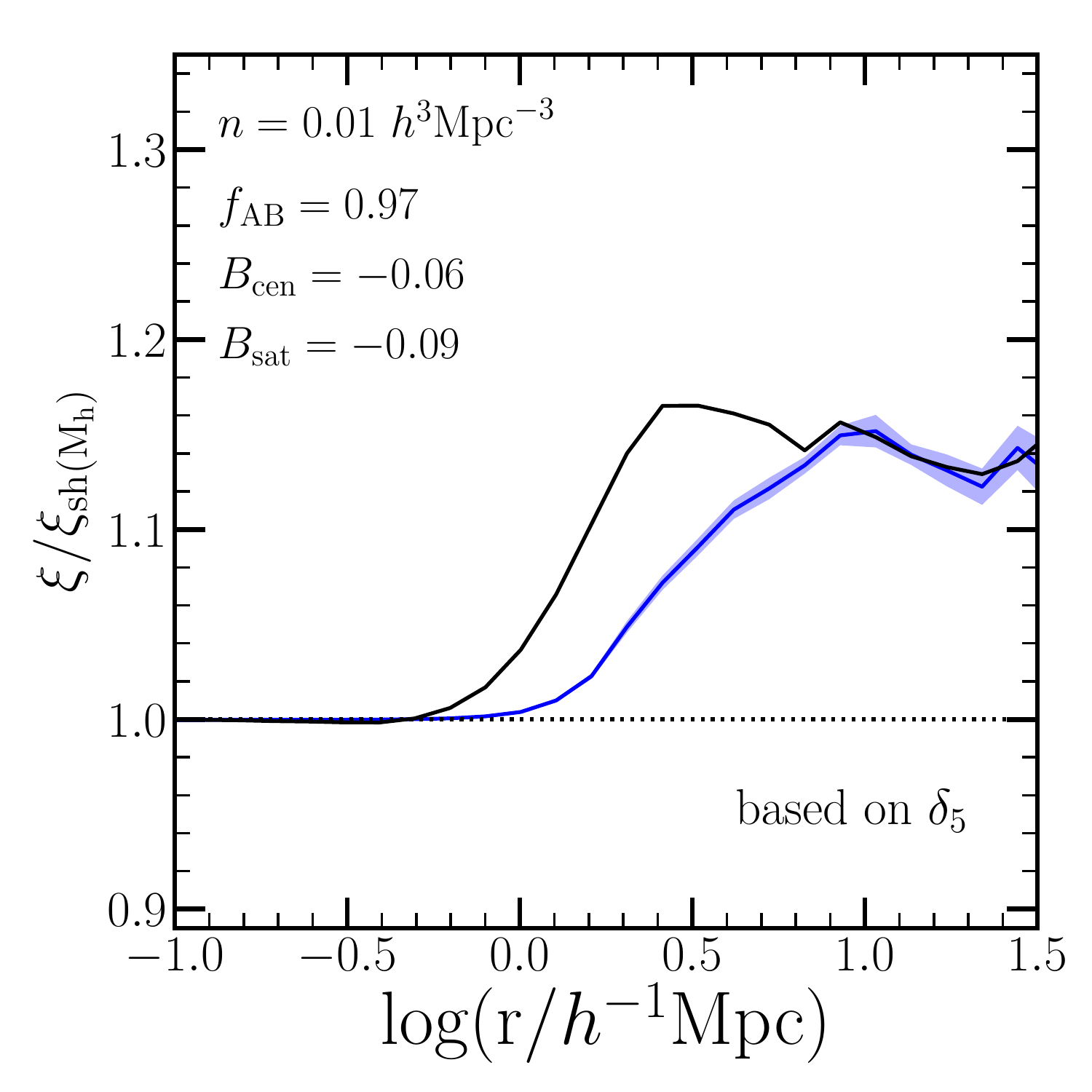}
	\end{subfigure}
	\hfill
\caption{
GAB level of modified HOD mock based on $\alpha_{\rm 0.3,1.25}$ (left) and $\delta_{\rm 5}$ (right), for the galaxy number density $n=0.01 \hmpcc$.
}
\label{fig:mHODgab_g5}
\end{figure}

In what follows, we explore the change in the HOD parameters as a function of ranked $\alpha_{\rm 0.3,1.25}$ and $\delta_{\rm 5}$ and assess further the applicability of our modified HOD model.

In \S~\ref{subsec:hodparam}, we study the changes in the standard HOD parameters as a function of ranked $\delta_{\rm 1.25}$, and find distinct changes for some of the parameters. Here we examine the same for $\alpha_{\rm 0.3, 1.25}$ in Fig.~\ref{fig:mHODmock_alpha03g125} and for $\delta_{\rm 5}$ in Fig.~\ref{fig:mHODmock_g5} for the $n=0.01 \hmpcc$ galaxy sample. The left-hand side of the figures shows the measured OV with these two properties in the SAM (red, blue, and black dots, as labelled). The six panels on the right-hand side of each figure show the values of the HOD parameters, fitted to the occupation functions of 10 per cent subsets of ranked environmental values (green points with errorbars). The horizontal lines in all subpanels represent the values of the HOD parameters obtained for the full sample.
We find similar dependences of the HOD parameters for both these environmental proxies. The centrals occupation parameters show a weak dependence on the ranked property, with ${\rm log} M_{\rm min}$ having an overall slightly decreasing trend. The satellites mass parameters also tend to decrease with increasing rank, resulting also in a decrease of $M_1/M_{\rm min}$. These changes are also similar to the parameters variation with $\delta_{\rm 1.25}$ shown in Figs.~\ref{fig:fithod_g125} and \ref{fig:mHODparam_g125}, but with weaker trends.

In \S~\ref{sec:edhod}, we propose a simple seven-parameter modified HOD model to include the OV of a secondary property, and use $\delta_{\rm 1.25}$ as an example for fitting the additional GAB parameters. We follow the same procedure here for fitting the two additional assembly bias parameters, $B_{\rm cen}$ and $B_{\rm sat}$,  separately for $\alpha_{\rm 0.3, 1.25}$ and for $\delta_{\rm 5}$.  In the process, we also create mock galaxy samples based on these 7-parameter modified HOD. The OV measured in the mock samples are represented by the curves shown on the left-hand side of Figs.~\ref{fig:mHODmock_alpha03g125} and Fig.~\ref{fig:mHODmock_g5}.  The modified HOD models selected are shown as the green lines in the six subpanels on the right-hand side of the figures.  We see that in both these cases, the modified HOD model is able to reasonably capture the change in the HOD parameters and produce the same level of OVs. The values of the assembly bias parameters are labelled in the figures. They all have negative values in accordance with the shift toward lower halo masses with increased value of the environmental property.  Their amplitudes are roughly comparable for $\alpha_{\rm 0.3, 1.25}$ and $\delta_{\rm 5}$, but somewhat smaller than the values found for $\delta_{\rm 1.25}$ in accordance with their weaker trends.

Finally, in Fig.~\ref{fig:mHODgab_g5}, we show the GAB level of the above mocks for $\alpha_{\rm 0.3,1.25}$ (left) and $\delta_{\rm 5}$ (right). In both cases, our seven-parameter HOD model is able to reproduce the correct level of GAB as in the SAM. We note that while the shuffling test indicated a larger level of GAB associated with $\delta_{\rm 5}$ (Fig.~\ref{fig:xienv}), our methodology is effectively able to ``tune down'' the assembly bias parameters to match the target level of GAB.  This indicates that our modified model is more general and extends beyond $\delta_{\rm 1.25}$ to other parameters that can represent the full level of GAB.


\bsp	
\label{lastpage}
\end{document}